\documentclass[10pt,journal,doublecolumn,twoside]{IEEEtran}
\usepackage{amsfonts}
\usepackage{}
\IEEEoverridecommandlockouts

\usepackage{amsthm}
\usepackage{subfigure}
\usepackage{colortbl}
\usepackage{bm}
\usepackage{soul}

\usepackage{graphicx}  
\usepackage{url}       
\usepackage{rotating}
\usepackage{amsmath}   
\usepackage{cite}

\usepackage{amsfonts,amssymb}

\usepackage{stfloats}

\usepackage{cases}
\usepackage{algorithm}
\usepackage{multirow}
\usepackage{algorithmic}
\usepackage{soul}
\usepackage{xcolor}
\usepackage{multirow}

\newcommand{\myvec}[1]%
{\stackrel{\raisebox{-2pt}[0pt][0pt]
{\small$\rightharpoonup$}}{#1}}

\newcommand{\ls}[1]
    {\dimen0=\fontdimen6\the\font
     \lineskip=#1\dimen0
     \advance\lineskip.5\fontdimen5\the\font
     \advance\lineskip-\dimen0
     \lineskiplimit=.9\lineskip
     \baselineskip=\lineskip
     \advance\baselineskip\dimen0
     \normallineskip\lineskip
     \normallineskiplimit\lineskiplimit
     \normalbaselineskip\baselineskip
     \ignorespaces
    }

\begin{document}
\setlength{\columnsep}{0.24in}

\title{
 Integrated Sensing and Communication for Anti-Jamming with OAM
 }

\author{Liping Liang,~\IEEEmembership{Member,~IEEE}, Wenchi Cheng,~\IEEEmembership{Senior Member,~IEEE}, Wei Zhang,~\IEEEmembership{Fellow,~IEEE}, \\
Zhuohui Yao,~\IEEEmembership{Member,~IEEE}


\thanks{\ls{.5}

Liping Liang, Wenchi Cheng, and Zhuohui Yao are with the State Key Laboratory of Integrated Services Networks, Xidian University, Xi'an, 710071, China (e-mail: liangliping\@xidian.edu.cn; wccheng\@xidian.edu.cn; yaozhuohui\@xidian.edu.cn).

Wei Zhang is with the School of Electrical Engineering and Telecommunications, the University of New South Wales, Sydney, Australia (e-mail: wzhang\@ee.unsw.edu.au).

}
}
\maketitle

\begin{abstract}

The spectrum share and open nature of wireless channels enable integrated sensing and communication (ISAC) susceptible to hostile jamming attacks, particularly in the context of partial band/broadband jamming attacks. Due to the intrinsic orthogonality and rich azimuth angle information of orbital angular momentum (OAM), vortex electromagnetic waves with helical phase fronts have shown great potential to achieve high-resolution imaging and strong anti-jamming capability of wireless communication. Focusing on significantly enhancing the anti-jamming results of ISAC systems with limited bandwidth under hostile jamming, in this paper we propose a novel ISAC for anti-jamming with OAM scheme, where the OAM legitimate transmitter can simultaneously sense the position of jammers with dynamic behavior and send data to multiple OAM legitimate users. Specifically, the OAM modes for sensing and communications are respectively hopped according to pre-set index modulation information to suppress jamming. To acquire the position of the jammer, we develop the enhanced multiple-signal-classification-based three-dimension position estimation scheme with continuous sensing in both frequency and angular domains, where the OAM transmitter is designed with the concentric uniform-circular-array mono-static method, to significantly increase the azimuthal resolution. Then, based on the acquired jamming channel state information, we develop the joint transmit-receive beamforming and power allocation scheme, where the transmit and receive beamforming matrices are dynamically adjusted to mitigate the mixed interference containing inter-mode interference, inter-user interference, and jamming, thus maximizing the achievable sum rates (ASRs) of all users. Numerical results demonstrate that our proposed scheme can significantly increase the ASR under broadband jamming attacks and achieve high detection accuracy of targets as compared with the conventional multiple-input-multiple-output (MIMO)-based ISAC.

\end{abstract}


\section{Introduction}

\IEEEPARstart{F}{uture} networks are expected to not only improve or extend the existing wireless communication technologies to satisfy various requirements of data rate or quality of service, but also provide various high-precision sensing services for intelligent applications, such as localization and tracking~\cite{10054381,xiao2022overview}. Integrated sensing and communications (ISAC), in which the sensing and wireless communication functionalities are integrated in a spectrum/hardware/energy-efficient way, recently has attracted much attention as a crucial technology to satisfy the requirements of both high-performance sensing and wireless communications for many application scenarios~\cite{9705498,9999292,xu2023robust}. 

However, ISAC is vulnerable to being attacked by hostile jammers and eavesdroppers due to the spectrum share and open nature of wireless channels in an increasingly complicated electromagnetic environment~\cite{9755276}. It is challenging for ISAC legitimate transceivers to distinguish jamming signals and desired signals, especially for partial band/broadband jamming signals, thus severely degrading the dual-functional quality of both communication and sensing. This motivates academics to explore efficient technologies for ISAC to overcome the physical-layer security issue~\cite{9927490,9973319,10143420,9681482,8445996,10499961}. Considering the scenarios where the eavesdropper acts as the target, the secure reflecting-intelligent-surface (RIS)-aided ISAC scheme was studied to design the phase shifts at the RIS and the communication-radar beamformers, thus maximizing the sensing beampattern gain~\cite{10143420}. The aerial RIS assisted ISAC scheme, where the aerial RIS was deployed to provide line-of-sight (LoS) links between the transmitter and targets, was proposed to maximize the achievable sum rate (ASR) while satisfying the requirement of the echo signal-to-interference-plus-noise ratio (SINR) for sensing with known jamming channel state information (CSI)~\cite{10464352}. The Stackelberg-game-based anti-jamming beamforming approach for ISAC systems was investigated with incomplete CSI~\cite{10543060}. Nevertheless, these works have several limits, such as perfect jamming CSI~\cite{10464352,8445996,10499961} and large bandwidth~\cite{9681482}. Until now, few works are related to anti-jamming technologies of ISAC systems.

Recently, the emerging orbital angular momentum (OAM), which enables the helical phase fronts of electromagnetic waves, has shown great potential for solving the critical physical-layer security problems of wireless communications due to the intrinsic orthogonality among different OAM modes~\cite{8336971,9690469,9354813,8712342,9145578,9844288,8558508,9791313,10040606,7797488}. The concept of mode hopping (MH) was proposed to resist malicious jamming attacks~\cite{8336971,9690469,9354813}, where one or several OAM-modes are dynamically activated at each hop according to pre-shared secrete keys between the legitimate transmitter and receivers. A secure range-dependent transmission scheme, where OAM-modes were generated by a frequency diverse array, was proposed to resist eavesdropping and obtain high energy efficiency by encoding the confidential information on OAM-modes~\cite{8712342}. The authors of~\cite{9145578} derived the secrecy capacity of uniform-circular-array (UCA) based OAM multiplexing wiretap systems. Moreover, the authors of~\cite{8558508,9791313,10040606,7797488} have demonstrated that OAM can achieve high spectrum efficiencies of wireless communications with lower complexity than multiple-input-multiple-output (MIMO) but consuming no extra resources, such as frequency, time, and power. By circumventing singular value decomposition operation, multiuser OAM-based Terahertz communication was studied to simplify the signal detection as compared with traditional MIMO systems~\cite{10040606}.

Apart from the above-mentioned advantages of OAM, it also posses several distinctive advantages in the field of position estimation, such as high azimuthal resolution without the necessity for beam scanning and accurate spinning velocity estimation in the absence of radial motion between targets and sensors~\cite{8653867,9968273}. With the assistance of OAM, the high-precision CSI between the transmitter and targets can be obtained. To eliminate the Bessel function modulation effect of OAM beams, a joint low-rank and sparse constraints-based radar imaging scheme was proposed, thus achieving high resolution~\cite{10260683}. Overall, OAM bridges a new way to solve the anti-jamming issue of ISAC systems under unknown jamming CSI without requiring additional frequency, time, power, and code resources.

Motivated by the above-mentioned discussion, in this paper we propose a novel ISAC for anti-jamming with OAM scheme, where the legitimate transmitter can simultaneously sense the position of the jammer and send data to multiple legitimate users at the same spectrum, to mitigate the interference by adjusting the transmit and receive beamforming, thus enhancing the anti-jamming results. In particular, the OAM modes for sensing and communications are hopped according to the pre-set index modulation information to suppress jamming.
We develop the enhanced multiple-signal-classification (EMUSIC) based three-dimension (3D) position estimation scheme with continuous sensing in both two-dimension (2D) frequency and angular domains to accurately estimate the position of the jammer, thus acquiring the jamming CSI. Then, we jointly design the transmit-receive (Tx-Rx) beamforming and power allocation scheme to mitigate the inter-mode, inter-user, and jamming interference. The optimization problem of maximizing the achievable sum rate (ASR) of all legitimate users is formulated, which is subject to the constraints of transmit power allocation, normalized power of Tx-Rx beamforming matrices, and the minimum sensing threshold. To achieve the maximum ASR, the joint Tx-Rx beamforming and power allocation alternating optimization (AO) design scheme is proposed. Numerical results show that the proposed ISAC for anti-jamming with OAM system can significantly increase the ASR under unknown and dynamic jamming as compared with existing systems.

The rest of this paper is organized as follows. Section~\ref{sec:sys} presents the system model of  ISAC for anti-jamming with multiuser OAM under hostile jamming. Section~\ref{sec:signal} gives the signal model, the relative position estimation between the jammer and transmitter, and the OAM communication signal extraction. In Section~\ref{sec:capa}, the joint Tx-Rx beamforming and power allocation scheme is proposed to maximize the ASR of all legitimate users for our proposed system. The performance of our proposed schemes is evaluated in Section~\ref{sec:Num} and the conclusions are presented in Section~\ref{sec:conc}.

{\emph{Notation}}: Bold uppercase and lowercase letters denote a matrix (i.e. $\bf{H}$) and a column vector (i.e. $\bf{x}$), respectively. $|\cdot|$, $\binom{a}{b}$, and $\lfloor\cdot\rfloor$ represent the absolute value of a number, the binomial coefficient by taking $b$ out of $a$, and the floor function, respectively. ${\bf H}\in \mathcal{C}^{N\times M}$ represents the matrix with the dimension of $N\times M$. Matrix superscripts $(\cdot)^{T}$, $(\cdot)^{*}$, $(\cdot)^{-1}$ and $(\cdot)^{H}$ denote the transpose, complex conjugate and transpose, Moore-Penrose inverse, and Hermitian transpose of a matrix, respectively. $\mathbb{E}(\cdot)$, $\odot$, $\circ$, and $\otimes$ represent the expectation operation, convolution, the Hadamard product, and the Kronecker product, respectively. ${\bf I}_{N}$ represents the identity matrix with the dimension of $N$. Also, $\parallel\cdot\parallel$ denotes the Frobenius norm of a vector. $[{\bf H}]_{n}$ represents the $n$-th column of matrix ${\bf H}$.

\begin{figure}[hb]
  \centering
  \includegraphics[width=0.485\textwidth]{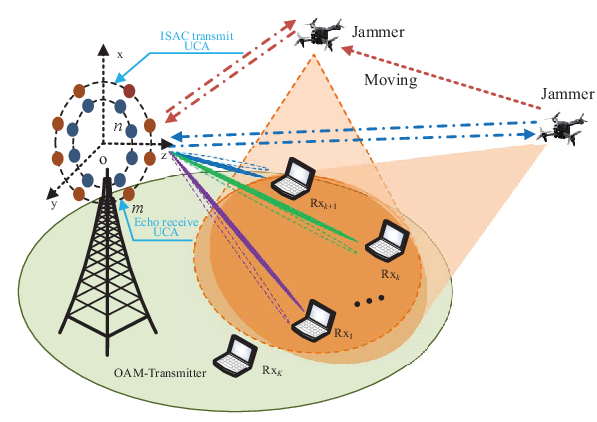}
  \caption{The architecture of ISAC for anti-jamming with OAM system.}
  \label{fig:sys}
\end{figure}

\section{System Model}\label{sec:sys}
As shown in Fig.~\ref{fig:sys}, we build up a new ISAC for anti-jamming with OAM system where the transmitter sends OAM-based ISAC signals simultaneously to $K$ legitimate OAM Rxs (also called users) for communications and a malicious $N_{\rm J}$-array jammer for sensing. The OAM transmitter and legitimate users are with $N_{t}$ arrays equidistantly distributed around the transmit and receive circles, respectively. The jammer attempts to send jamming signals to interrupt the communications between the OAM transmitter and legitimate OAM users. Without loss of generality, jamming signals are sent with plane electromagnetic waves which can be considered as vortex electromagnetic waves with zero OAM mode. As shown in Fig.~\ref{fig:sys}, the jammer dynamically moves at a low speed near the legitimate OAM users. Thus, the jamming converge area varies with the movement of the jammer. In practical scenarios, it is very hard to keep strict alignment between the jammer and legitimate receive UCAs. Thus, the inter-mode interference caused by jamming to legitimate users varies with the relative locations between the jammer and legitimate OAM users. For instance, when the malicious jammer moves next to the $k$-th ($1 \leq k \leq K$) OAM user, denoted by ${\rm{Rx}}_{k}$, the interference to ${\rm{Rx}}_{k}$ increases. If ${\rm{Rx}}_{k}$ having no information about the jammer movement trajectory still utilizes the previous beamforming matrix and power allocation scheme, the interference increases. Inversely, when the jammer moves departure from ${\rm{Rx}}_{k}$ but next to the $(k+1)$-th OAM user, denoted by ${\rm{Rx}}_{k+1}$, ${\rm{Rx}}_{k}$ and ${\rm{Rx}}_{k+1}$ are required to adjust their beamforming matrices to mitigate the interference. Therefore, the non-time-real beamforming and power allocation degrades the ASR of all legitimate OAM receivers in ISAC for anti-jamming with OAM systems.

\begin{figure}[hb]
  \centering
  \includegraphics[width=0.485\textwidth]{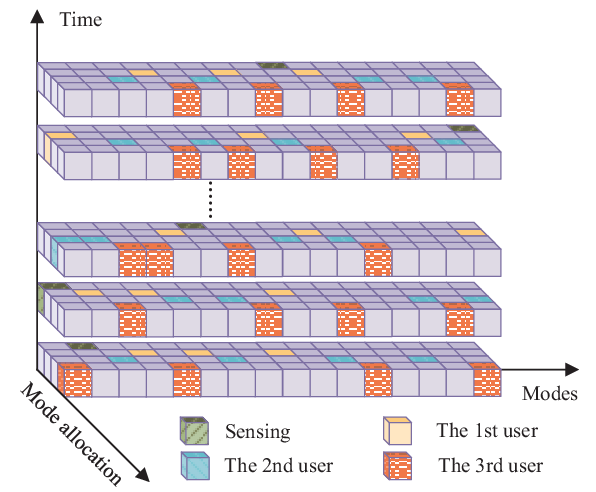}
  \caption{The index-modulation-based mode hopping pattern of our proposed systems.}
  \label{fig:OAM_ISAC_MH}
  \vspace{-10pt}
\end{figure}

To solve the problem of jammer location information acquisition, the OAM transmitter simultaneously integrates sensing to the malicious jammer and communications to all legitimate OAM users through LoS links in the proposed ISAC system as depicted in Fig.~\ref{fig:sys}. The OAM transmitter divides OAM beams into two parts: one points to legitimate OAM users for data transmission and the other one points to the jammer for position detection. Also, the OAM transmitter, which is design as concentric UCAs as shown in Fig.~\ref{fig:sys}, can receive the reflected echo OAM signals from the jammer. The deign of OAM transmitter is mono-static OAM-MIMO sensing. The distance and azimuth information about the jammer are obtained in both frequency and angular dimensions. In the proposed ISAC system, the OAM modes for ISAC are determined by the input index modulation which is pre-shared by the OAM transmitter and each legitimated OAM users. For efficient anti-jamming, the allocated OAM modes for communication vary according to the pre-shared index modulation as shown in Fig.~\ref{fig:OAM_ISAC_MH}. Therefore, the previous and the next symbols are probably carried by different OAM modes to legitimate OAM users.


\section{Jammer Position Estimation \\ and Interference Mitigation}\label{sec:signal}
To mitigate the impact of jamming on legitimate signals of multiple OAM users (also called receivers), the position of the jammer is first detected. According to the detection results, the OAM transmitter controls the beamforming matrices and changes the allocated power for multiuser wireless communications, thus resulting in increasing the ASR of ISAC for anti-jamming with OAM systems.

\subsection{Transmitter Design}\label{subsec:TX}

To generate multiple OAM modes, the $N_{t}$ transmit arrays are fed with the same input modulated signals along with a successive phase array $e^{j\frac{2\pi l}{N_{t}}}$ from array to array, where $l$ satisfying $|l| \leq \frac{N_{t}}{2}$ represents the order of OAM modes. The generated multiple OAM modes can be used simultaneously over the same frequency band at the same time. We assume that $N_{f}$ orthogonal subcarriers are used in the proposed ISAC system. As mentioned in Section~\ref{sec:sys}, the available $N_{t}$ OAM modes are divided into $(K+1)$ non-overlapping parts to simultaneously sense the malicious jammer and transmit data to $K$ legitimate OAM users, where the $k$-th legitimate user is assigned to $ N_{k} \geq 1 $ OAM modes. The OAM set assigned to the $k$-th legitimate user is defined by $\mathcal L_{k}=\{l_{1_{k}},\cdots,l_{i_{k}}\cdots,l_{N_{k}}\}$, where $l_{i_{k}} (i_{k} \in [1, N_{k}])$ is the order of the $i$-th OAM mode for the $k$-th user.

To obtain much information about the jammer position, all the modes are activated one by one within a sensing duration $T_{s}$. 
The $N_{t}$ OAM modes are jointly used with $N_{ f}$ subcarriers within each orthogonal frequency division multiplexing (OFDM) symbol duration, denoted by $T$, in the proposed ISAC systems. We denote by $l_{i_{s}}$ and $\Delta f=B/N_{ f}=1/T$ the OAM mode for sensing and the subcarrier spacing with $B$ being the bandwidth, respectively. Thus, we have $T_{s}=N_{t}(T+T_{\rm cp})$, where $T_{\rm cp}$ is the cyclic prefix transmission time. Therefore, the transmit ISAC signal, denoted by ${\bf x}_{q}$, over the $q$-th ($q=0,1,\cdots,N_{ f}-1$) subcarrier for the OAM system is expressed as follows:
\begin{equation}
{\bf x}_{q} = {\bf x}_{q,{ c}} + {\bf x}_{q,{s}},
\end{equation}
where ${\bf x}_{q,{\rm c}} \in \mathcal{C}^{N_{t}\times 1}$ and ${\bf x}_{q,{\rm s}}\in \mathcal{C}^{N_{t}\times 1}$ representing the transmit communication and sensing signals, respectively, over the $q$-th subcarrier are given by
\begin{equation}
  \begin{cases}
 {\bf{x}}_{q,c} = \sum\limits_{k=1}^{K}{\bf{F}}^{ H}{\bf P}_{kq}{\bf{s}}_{q,k};
\\
{\bf x}_{q,s}=  {\bf {F}}^{  H} {\bf P}_{sq} {\bf{ s}}_{q,{\rm s}}.
  \end{cases}
\label{eq:xq}
\end{equation}

\setcounter{equation}{11}
\begin{figure*}[!hb]
\hrulefill
\begin{eqnarray}
h_{gq,i_{s}}\!\!\!\!\! &=& \!\!\!\!\!\frac{ e^{-j2\pi f_{0} \tau(t,g)} e^{j 2\pi q \Delta f (t-\tau(t,g))}}{ N_{t}}
\sum_{m=1}^{N_{t}}\sum_{n=1}^{N_{t}}h_{mn,gq} e^{j\frac{2\pi n l_{i_{s}}}{N_{t}}} 
\nonumber\\
 \!\!\!\!\!& = &\!\!\!\!\! \frac{\beta \lambda_{q} \chi_{g} e^{-\!j2\pi f_{0} \tau(t,g)} e^{j 2\pi q \Delta f (t\!-\!\tau(t,g))}
 e^{\!-j\frac{2\pi}{\lambda_{q}}\left(\sqrt{R_{g}^{2}+\!r_{t}^{2}}+\sqrt{R_{g}^{2}+r_{r}^{2}}\right)}}{4\pi N_{t}\sqrt{\left(R_{g}^{2}+\!r_{t}^{2}\right)\left(R_{g}^{2}+\!r_{r}^{2}\right)}}
 \!\!\!\sum_{m=1}^{N_{t}}\! e^{j\frac{2\pi r_{r}R_{g}\sin\phi_{g}\cos \theta_{mg}}{\lambda \sqrt{R_{g}^{2}+r_{r}^{2}} }  } 
\!\! \sum_{n=1}^{N_{t}} \! e^{j\frac{2\pi r_{t}R_{g}\sin\phi_{g}\cos\theta_{ng}}{\lambda \sqrt{R_{g}^{2}+r_{t}^{2}} }} e^{j (\theta_{g}\!-\!\theta_{ng}) l_{i_{s}} }
 \nonumber\\
\!\!\!\!\!  & = & \!\!\! \!\!j^{-2 l_{i_{s}}} A_{q,i_{s},g}\chi_{g}  e^{-j2\pi f_{0} \tau_{g}} e^{j2\pi f_{d}t}  e^{j 2\pi q \Delta f (t-\tau_{g}+\frac{2 v t}{c})},
   \label{eq:hls}
\end{eqnarray}
\vspace{-10pt}
\end{figure*}
 In Eq.~\eqref{eq:xq}, ${\bf F}=[{\bf f}_{1},{\bf f}_{2},\cdots,{\bf f}_{n},\cdots,{\bf f}_{N_{t}}]$ is the discrete Fourier transform (DFT) matrix with entries $\frac{e^{-j\frac{2\pi (n-1) l}{N_{t}}}}{\sqrt{N_{t}}}$ in the $n$-th ($n=1,2,\cdots,N_{t}$) column and the $(l+\lfloor\frac{N_{t}}{2}\rfloor)$-th row. ${\bf P}_{sq}={\rm diag}\{\sqrt{P_{qi_{s}}}\}$ and ${\bf P}_{kq} = {\rm diag}\{\sqrt{P_{qi_{k}}}\}$ are the transmit power diagonal matrices for sensing and communications, respectively. It is noticed that the entry in ${\bf P}_{sq}$ or ${\bf P}_{kq}$ corresponding to the OAM mode $l_{i_{j}}$ ($i_{j} = \{i_{k},i_{s}\}$) equals $\sqrt{P_{qi_{j}}}$, while the remaining entries are zeros. Denoting the total transmit power by $P_{t}$ for sensing and communication, we thus have
 \setcounter{equation}{2}
 \begin{equation}
    {\rm Tr} \left(\sum_{q=1}^{N_{f}}{\bf P}_{sq}{\bf P}_{sq}^{T} +\sum_{k=1}^{K}\sum_{q=1}^{N_{f}}{\bf P}_{kq}{\bf P}_{kq}^{T}\right) \leq  P_{t}
    \label{eq:power_sum}
\end{equation}
 within the sensing duration $T_{s}$. In addition, ${\bf s}_{q,{\rm s}} \in \mathcal{C}^{N_{t}\times 1}$ and ${\bf s}_{q,k} \in \mathcal{C}^{N_{t}\times 1}$ denote the transmit sensing and communication signals over the $q$-th subcarrier expressed by
 \begin{equation}
  \begin{cases}
  {\bf s}_{q,{ s}}=\frac{e^{j2\pi f_{q}t}}{\sqrt{N_{ f}}}s_{qi_{s}}{\bf e}_{l_{i_{s}}+\lfloor\frac{N_{t}}{2}\rfloor};
\\
{\bf s}_{q,k}=  \frac{e^{j2\pi f_{q}t}}{\sqrt{N_{ f}}}\sum\limits_{i_{k}=1}^{N_{k}}s_{qi_{k}}{\bf e}_{l_{i_{k}}+\lfloor\frac{N_{t}}{2}\rfloor},
  \end{cases}
\end{equation}
 where $t$ is the time variable, $s_{qi_{j}}$ is denoted by the transmit signals on the OAM mode $l_{i_{j}}$ at the $q$-th ($q=0,1,\cdots,N_{ f}-1$) subcarrier, and ${\bf e}_{l_{i_{j}}}$ is the unit vector with the entry equal to 1 at the $\left(l_{i_{j}}+\lfloor\frac{N_{t}}{2}\rfloor\right)$-th position. For simplicity, ${\bf e}_{l_{i_{j}}+\lfloor\frac{N_{t}}{2}\rfloor}$ is represented by ${\bf e}_{i_{j}}$ in the following.

\subsection{Jammer Sensing in 2D Domains}\label{subsec:sensing}

To obtain the position information of the jammer, the received OAM echo signals from the jammer are required. According to~\cite{8630982} and~\cite{8960405}, the initial order of OAM modes is changed into the opposite one after OAM beams first-order reflection. Hence, the reflected sensing signal of ${\bf x}_{q,{ s}}$, denoted by $\tilde{\bf x}_{q,{ s}}$, over the $q$-th subcarrier within the sensing duration $T_{s}$ can be expressed as follows:
\begin{eqnarray}
 \tilde{\bf x}_{q,{ s}}= \frac{e^{j2\pi f_{q}t}}{\sqrt{N_{ f}}}\sum\limits_{i_{s}=1}^{N_{t}}{\bf {F}}^{ H} {\bf P}_{sq} s_{qi_{s}}{\bf e}_{-i_{s}}.
 \label{eq:fs}
\end{eqnarray}

Assuming that the jammer is comprised of $G$ reflect points, we thus have the received echo signal, denoted by ${\bf y}_{gq} \in \mathcal{C}^{N_{t}\times 1}$, from the $g$-th ($g=1,2,\cdots, G$) reflection point at the $q$-th subcarrier as follows:
\begin{equation}
  {\bf y}_{gq}={\bf H}_{gq} \odot \tilde{\bf x}_{q,s}+{\bf n}_{gq},
  \label{eq:yqg}
\end{equation}
where ${\bf{n}}_{gq} $ is the received additional white Gaussian noise (AWGN) with zero means and ${\bf H}_{gq}$ is the channel matrix, respectively, corresponding to the $g$-th reflection point at the $q$-th subcarrier.


To derive the channel gain ${\bf H}_{gq}$, we assume $(R_{g}, \phi_{g},\theta_{g})$ the spherical coordinate of the $g$-th reflection point $P_{g}$ to the center of OAM transmitter, where $R_{g}$, $\theta_{g}$, and $\phi_{g}$ are the distance, azimuth angle, and elevation angle, respectively. As illustrated in Fig.~\ref{fig:sys}, the transmit and echo receive UCAs are located on the x-y plane. The spherical coordinates of the $n$-th transmit array over the UCA with the radius $r_{t}$ and the $m$-th ($m=1,2,\cdots,N_{t}$) receive array over the UCA with the radius $r_{r}$ are $\left(r_{t},0,\frac{2\pi n}{N_{t}}\right)$ and $\left(r_{r},0,\frac{2\pi m}{N_{t}}\right)$, respectively. We denote by $d_{gn}$ and $d_{mg}$ are the distances from the $n$-th transmit array to the $g$-th reflect point and from the $g$-th reflect point to the $m$-th receive array, respectively. Thus, we have the channel gain, denoted by $h_{mn,gq}$, from the $n$-th transmit array to the $m$-th receive array as follows:
\begin{equation}
 h_{mn,gq}=\frac{\beta \lambda_{q} \chi_{g}}{4\pi d_{gn} d_{mg}}e^{-j\frac{2\pi (d_{gn}+d_{mg})}{\lambda_{q}}} \delta(t-\tau(t,g)),
\end{equation}
where $\lambda_{q}$ denotes the wavelength of the frequency $f_{q}$, $\beta$ represents the attenuation constant, $\chi_{g}$ represents the radar cross section of $g$-th reflection point, $\tau(t,g)$ denotes the time delay through the sensing link between the ISAC transmitter and the $g$-th reflect point, and $\delta(t)$ is the impulse response. $h_{mn,gq}$ is the entry in the $m$-th row and $n$-th column of ${\bf H}_{gq}$.


Based on the Taylor series expansion and $r_{t}, r_{r}$ are far smaller than the distance between the centers of ISAC UCAs and the jammer, $d_{gn}$ and $d_{mg}$ can be approximately expressed as follows:
\begin{subnumcases}
\!\!\!\!d_{gn}\!\!\approx \!\sqrt{R_{g}^{2}+r_{t}^{2}}- \!\frac{r_{t}R_{g}\sin\phi_{g}\cos\left(\frac{2\pi n}{N_{t}}- \!\theta_{g}\right)}{\sqrt{R_{g}^{2}+r_{t}^{2}}};
\label{eq:dns}\\
 \!\!\!\!d_{mg}\!\!\approx \!\sqrt{R_{g}^{2}+r_{r}^{2}}- \!\frac{r_{r}R_{g}\sin\phi_{g}\cos\left(\frac{2\pi m}{N_{t}}- \!\theta_{g}\right)}{\sqrt{R_{g}^{2}+r_{r}^{2}}}.
 \label{eq:dms}
\end{subnumcases}
Thus, we can use the first terms on the right hand of Eqs.~\eqref{eq:dns} and \eqref{eq:dms} as amplitudes and the second terms as phases for $d_{gn}$ and $d_{mg}$, respectively. Hence, $h_{mn,gq}$ can be derived as follows:
\begin{eqnarray}
h_{mn,gq} \hspace{-0.3cm}&=& \hspace{-0.3cm}e^{j\frac{2\pi}{\lambda_{q}}\left(\frac{r_{t}R_{g}\sin\phi_{g}\cos\left(\frac{2\pi n}{N_{t}}-\theta_{g}\right)}{\sqrt{R_{g}^{2}+r_{t}^{2}}}
+\frac{r_{r}R_{g}\sin\phi_{g}\cos\left(\frac{2\pi m}{N_{t}}-\theta_{g}\right)}{\sqrt{R_{g}^{2}+r_{r}^{2}}}\right)}
\nonumber\\
&& \times \frac{\beta \lambda_{q} \chi_{g} \delta(t-\tau(t,g)) e^{-j\frac{2\pi}{\lambda_{q}}\left(\sqrt{R_{g}^{2}+r_{t}^{2}}+\sqrt{R_{g}^{2}+r_{r}^{2}}\right)}}{4\pi \sqrt{\left(R_{g}^{2}+r_{t}^{2}\right)\left(R_{g}^{2}+r_{r}^{2}\right)}},
\nonumber\\
  \label{eq:hgmn}
\end{eqnarray}
where given the light speed $c$ and radial velocity $v$ between the jammer and the ISAC transmitter, we have
\begin{equation}
    \tau(t,g)=\frac{2(R_{g}-vt)}{c}.
\end{equation}

Based on Eqs.~\eqref{eq:fs} and~\eqref{eq:yqg}, we have the decomposed OAM signals, denoted by $ \tilde{\bm y}_{gq}$, at the $q$-th subcarrier within the sensing duration $T_{s}$ as follows:
\begin{eqnarray}
  \tilde{\bm y}_{gq}= {\bf E}^{ H}{\bf H}_{gq}\otimes {\bf {F}} {\bf P}_{sq}{\bf{ s}}_{q,{\rm s}}+{\bf E}^{ H}{\bf n}_{gq},
  \label{eq:ysqg}
\end{eqnarray}
where the entries of ${\bf E} \in \mathcal{C}^{N_{t}\times N_{t}}$ are one.

Substituting Eq.~\eqref{eq:hgmn} into Eq.~\eqref{eq:ysqg}, we can derive the channel gain, denoted by $h_{gq,i_{s}}$, of the first term on the right hand of Eq.~\eqref{eq:ysqg} with respect to $s_{qi_{s}}$ after down-conversion processing as shown in Eq.~\eqref{eq:hls},
where $J_{l}(z)$ given by
\setcounter{equation}{12}
\begin{equation}
  J_{l}(z)=\frac{j^{l}}{2\pi}\int_{0}^{2\pi}e^{j(z\cos\theta_{ng}-l\theta_{ng})} d\theta_{ng}
\end{equation}
is the $l$-th order Bessel function of the first kind, $f_{d}=\frac{2v f_{0}}{c}$ is the Doppler frequency shift, $\tau_{g}=\frac{2 R_{g}}{c}$, and
\begin{eqnarray}
  \!\!\!A_{q,i_{s},g}\!\!\!\!\!&=&\!\!\!\!\!\! \frac{\beta \lambda_{q}  N_{t}  e^{-j\frac{2\pi}{\lambda_{q}}\left(\sqrt{R_{g}^{2}+r_{t}^{2}}+\sqrt{R_{g}^{2}+r_{r}^{2}}\right)}
  e^{j 2\theta_{g} l_{i_{s}} }  }
  {4\pi \sqrt{\left(R_{g}^{2}+r_{t}^{2}\right)\left(R_{g}^{2}+r_{r}^{2}\right)}}
  \nonumber\\
 && \!\!\!\!\times J_{l_{i_{s}}}\!\!\!\left(\!\!\frac{2\pi r_{t}R_{g}\sin\phi_{g}}{\lambda_{q} \sqrt{R_{g}^{2}+r_{t}^{2}}}\!\!\right)
  J_{0}\!\!\left(\!\!\frac{2\pi r_{r}R_{g}\sin\phi_{g}}{\lambda_{q} \sqrt{R_{g}^{2}+r_{r}^{2}}}\!\!\right).
\label{eq:Aqisg}
\end{eqnarray}
Compensating the OAM mode independent term $e^{-2j l_{i_{s}} }$, we have the channel gain, denoted by $\tilde{h}_{gq,i_{s} }$, for the OAM mode $l_{i_{s}}$ as follows:
\begin{eqnarray}
\tilde{h}_{gq,i_{s} }= A_{q,i_{s},g} \chi_{g}e^{-j2\pi f_{0} \tau_{g}} e^{j2\pi f_{d}t}  e^{j2\pi q \Delta f (t-\tau_{g}+\frac{2 v t}{c})}.
\label{eq:hgq}
\end{eqnarray}
Sampling time set as $t+i_{s}T_{0}$ with $t=\frac{\ddot{q}}{\Delta f N_{ f}^{\prime}}$ and $T_{0}=T+T_{\rm cp}$ corresponding to the OAM mode $l_{i_{s}}$, the discrete-time vision of $\tilde{h}_{gq,i_{s} }$ can be expressed as follows:
\begin{eqnarray}
\!\!\!\!\!\!\!\tilde{h}_{gq,i_{s} }(\ddot{q})\!\!\! \!\!&=&\!\!\!\!\! A_{q,i_{s},g}  \chi_{g} e^{-j2\pi f_{0} \tau_{g}}
e^{j \frac{2\pi f_{d} \ddot{q} }{ N_{f} \Delta f }}  e^{j2\pi f_{d} i_{s} T_{0}}
\nonumber\\
&&\!\!\!\!
e^{j 2\pi q \Delta f  \left(\frac{\ddot{q}}{N_{f}^{\prime} \Delta f } +i_{s}T_{0} -\tau_{g}
+\frac{2v}{c}\left(\frac{\ddot{q}}{N_{f}^{\prime} \Delta f} + i_{s} T_{0}\right)\right)},
\label{eq:hqisq}
\end{eqnarray}
where $\ddot{q}=0,1,\cdots, N_{\rm f}^{\prime}-1$ and $N_{f}^{\prime} \geq N_{f}$. With $N_{f}^{\prime}$, the precision of the estimation can be enhanced but has no effect on the range resolution. To mitigate the OFDM phase effect of Doppler shifts, it is required that $T_{\rm cp} \geq max\{\tau_{g}\}$ and $v \ll \frac{c \Delta f}{ 2 f_{0}}$~\cite{10036975}.
Thus, $\left(e^{j\frac{2\pi f_{d} \ddot{q} }{N_{f} \Delta f}}, e^{j 2 \pi q \Delta f \left(\frac{2v}{c}\left(\frac{\ddot{q}}{N_{f} \Delta f} + i_{s} T_{0}\right)\right)}\right)$ approximates to $(1,1)$. After removing CP at the echo OAM receiver, we can obtain the $G$ received echo signals, denoted by $y_{qi_{s}}$, for the OAM mode $l_{i_{s}}$ at the $q$-th subcarrier as follows:
\begin{eqnarray}
    y_{qi_{s}}
& = &  \sqrt{P_{qi_{s}}} h_{q i_{s}}s_{qi_{s}}+n_{q i_{s}} ,
 \label{eq:yql}
\end{eqnarray}
where $n_{q i_{s}}$ is the received noise with respect to the OAM mode $l_{i_{s}}$ at the $q$-th subcarrier and $h_{q i_{s}}$ is given by
\begin{eqnarray}
h_{qi_{s}}  = \sum_{g=1}^{G} A_{q,i_{s},g}\chi_{g}  e^{-j 2 \pi f_{0} \tau_{g}} e^{j2\pi f_{d} i_{s} T_{0}}
e^{-j 2\pi q \Delta f \tau_{g}}.
\label{eq:hqisq}
\end{eqnarray}
Thus, the received sensing signal-to-noise ratio (SSNR) can be calculated. Based on Eq.~\eqref{eq:yql}, 
stacking ${\bf H}_{s}$, ${\bf S}_{s}$, and ${\bf N}_{s}$ by $h_{q i_{s}}$, $\sqrt{P_{qi_{s}}}s_{qi_{s}}$, and $n_{q i_{s}}$, respectively, with the dimension of $N_{f}\times N_{t}$, we have
\begin{equation}
  {\bf Y}_{s}={\bf H}_{s}\circ{\bf S}_{s}+{\bf N}_{s},
  \label{eq:hsq}
\end{equation}
By virtue of the shared transmit sensing signal ${\bf S}_{s}$ between the transmit and receive UCAs, the complex-value element-wise division of the received echo signal ${\bf Y}_{s}$ over ${\bf S}_{s}$ is expressed as follows~\cite{9282206,9269381}:
\begin{equation}
  \frac{{\bf Y}_{s}}{{\bf S}_{s}} ={\bf H}_{s}+\frac{{\bf N}_{s}}{{\bf S}_{s}}.
  \label{eq:Ddiv}
\end{equation}

\subsection{Position Acquisition of the Jammer}\label{subsec:position}
Based on Eq.~\eqref{eq:Ddiv}, we develop the EMUSIC algorithm to estimate the 3-D position of the jammer. In particular, the relative distance $R_{g}$, the relative velocity $v$, and the elevation angle $\phi_{g}$ can be estimated in the frequency domain. Meanwhile, the relative azimuth angle $\theta_{g}$ and the elevation angle $\phi_{g}$ between the ISAC transmitter and the jammer can be acquired in the OAM domain.



{\bf 1) $\phi_{g}$ and $\theta_{g}$ estimation in the OAM domain:} 
According to Eq.~\eqref{eq:hqisq}, the corresponding vector, denoted by ${\bf h}_{q} \in \mathcal{C}^{ N_{t} \times 1}$, of ${\bf H}_{s}$ consisting of the information about $\phi_{g}$ and $\theta_{g}$ for the $q$-th subcarrier can be expressed as follows:
\begin{equation}
    {\bf h}_{q} = {\bf A}_{q}{\boldsymbol{\chi}}+{\bf n}_{q},
\end{equation}
where ${\boldsymbol{\chi}}=[\chi_{1},\chi_{2},\cdots,\chi_{G}]^{T}$ is considered as the signals from reflection points, ${\bf n}_{q}$ is the noise vector, and 
\begin{equation}
    {\bf A}_{q}=\left[
    \begin{matrix}
      A_{q,1,1} & A_{q,1,2} & \cdots & A_{q,1,G}  \\
      A_{q,2,1} & A_{q,2,2} & \cdots & A_{q,2,G}  \\
        \vdots & \vdots   &  \ddots & \vdots   \\
      A_{q,N,1} & A_{q,N,2} & \cdots & A_{q,N,G}  \\
    \end{matrix}
    \right]
\end{equation}
is the direction matrix containing azimuth as well as elevation angle information of all reflection points at the $q$-th subcarrier.
Thus, the covariance matrix of ${\bf h}_{q}$ can be expressed as follows:
\begin{eqnarray}
    \begin{aligned}
      {\bf R}_{{\bf h}_{q}}& ={\bf A}_{q}\mathbb{E} \{{\bf \chi}{\bf \chi}^{ H}\}  {\bf A}_{q}^{ H} +\mathbb{E} \{{\bf n}_{q} {\bf n}_{q}^{ H}\}
      \nonumber\\
      & ={\bf A}_{q}{\bf \Sigma}_{\bf \chi}  {\bf A}_{q}^{ H} +{\bf \Sigma}_{{\bf n}_{q}} ,
    \end{aligned}
\end{eqnarray}
where ${\bf \Sigma}_{{\bf n}_{q}} = {\rm diag}\{\sigma^{2}_{qi_{s}}\}$ is the variance matrix of ${\bf n}_{q}$ and $\mathbb{E} \{{\bf\chi}_{q}{\bf \chi}_{q}^{\rm H}\}={\bf \Sigma}_{\bf \chi}$ is a diagonal matrix due to the uncorrelation of echo signals. Thus, we have ${\rm rank}({\bf \Sigma}_{\bf \chi}) =G $. Due to ${\rm rank}({\bf A}_{q})=G$, we can obtain ${\rm rank} \left({\bf A}_{q}  {\bf \Sigma}_{\bf \chi} {\bf A}_{q}^{ H} \right)=G$ with the assumption $G \leq N_{t}$.
In practice, the covariance matrix ${\bf R}_{{\bf h}_{q}}$ is calculated within $\mathcal{N}$ sampling sets of ${\bf h}_{q,{\tilde n}}$ expressed by~\cite{9173826,17564}
\begin{equation}
    {\bf R}_{{\bf h}_{q}}
     =  \frac{1}{\mathcal{N}}\sum_{\tilde{n}=1}^{\mathcal{N}} {\bf h}_{q,\tilde{n}} {\bf h}_{q,\tilde{n}}^{ H}.
\label{eq:cova_hq}
\end{equation}

With eigenvalue decomposition, ${\bf R}_{{\bf h}_{q}}$ can be re-expressed as follows:
\begin{equation}
    {\bf R}_{{\bf h}_{q}}= {\bf U}_{q} {\bf \Omega}_{q} {\bf U}_{q}^{ H}= \sum_{i_{s}=1}^{N_{t}}\mu_{q,i_{s}} {\bf u}_{q,i_{s}} {\bf u}_{q,i_{s}}^{ H},
    \label{eq:cova_evd}
\end{equation}
where $\mu_{q,1} \geq \mu_{q,2} \cdots \geq \mu_{q,N_{t}} $ are the eigenvalue of ${\bf R}_{{\bf h}_{q}}$, ${\bf \Omega}_{q}={\rm diag}\{\mu_{q,1},\cdots,\mu_{q,N_{t}}\}$ is the diagonal matrix, and ${\bf U}_{q}=[{\bf u}_{q,1}, {\bf u}_{q,2},\cdots,{\bf u}_{q,N_{t}}]$ is the eigen matrix composing of eigen vectors ${\bf u}_{q,i_{s}}$ with respect to $\mu_{q,i_{s}}$.

Assuming that the number of targets is estimated to be $\widehat{G}$, ${\bf R}_{{\bf h}_{q}}$ thus can be re-expressed by
\begin{equation}
{\bf R}_{{\bf h}_{q}}  = {\bf U}_{\chi_{q}} {\bf \Omega}_{{\bf \chi}_q} {\bf U}_{\chi_{q}}^{ H}
    + {\bf U}_{n_{q}}{\bf \Omega}_{ n_{q}} {\bf U}_{n_{q}}^{ H},
\end{equation}
where $ {\bf \Omega}_{{\bf \chi}_q}={\rm diag}\{\mu_{q,1},\cdots,\mu_{q,\widehat{G}}\}$ and ${\bf \Omega}_{ n_{q}}$ are diagonal matrices composed of the largest $\widehat{G}$ and the smallest $(N_{t}-\widehat{G})$ eigenvalues, respectively. Also, we have the matrices ${\bf U}_{\chi_{q}} = \left[{\bf u}_{q,1},\cdots,{\bf u}_{q,\widehat{G}}\right]$ regard as the signal space and ${\bf U}_{n_{q}} = \left[{\bf u}_{q,\widehat{G}+1},\cdots,{\bf u}_{q,N_{t}}\right]$ forming the noise space, which are composed of eigenvectors corresponding to the largest $\widehat{G}$ and the smallest $(N_{t}-\widehat{G})$ eigenvalues, respectively.

If $\widehat{G} < G$, the $(G-\widehat{G})$ reflection points are missed. When the jammer is far away from the OAM transmitter while the scattering clutters are near to, the OAM transmitter with traditional MUSIC algorithms would not detect the position of the jammer, thus causing ineffective anti-jamming results. To significantly increase the estimation accuracy, we reconstruct ${\bf U}_{n_{q}}$ as follows:
\begin{equation}
\widehat{\bf U}_{n_{q}} = \left[\hat{\bf u}_{q,\widehat{G}+1},\cdots,\hat{\bf u}_{q,G},\hat{\bf u}_{q,G+1},\cdots,\hat{\bf n}_{q,N_{t}}\right],
\end{equation}
where we define
\begin{equation}
\hat{\bf u}_{q,\widehat{G}+\kappa} = \left(\frac{ \rho \mu_{q,N_{t}}}{\mu_{q,\widehat{G}+\kappa}}\right)^{\nu} {\bf u}_{q,\widehat{G}+\kappa}
\label{eq: hat_u}
\end{equation}
with the coefficients $\rho$, $\nu$, and $\kappa  = \{1,2,\cdots,N_{t}-\widehat{G}\}$.

Based on Eq.~\eqref{eq:Aqisg}, we have the direction vector ${\bf a}_{q}=\left[a_{q1},\cdots,a_{qi_{s}},\cdots,a_{qN_{t}}\right]^{ T}$ for the $q$-th subcarrier with the $i_{s}$-th entry given by
\begin{equation}
    a_{qi_{s}} = e^{2j\theta l_{i_{s}}}J_{l_{i_{s}}}\left(\frac{2\pi r_{t}\sin\phi } {\lambda_{q} }\right) ,
\end{equation}
where $\theta \in [0,2\pi]$ and $\phi \in [0,\pi]$ are the azimuth and elevation angles, respectively.

Therefore, the spatial-spectrum function, denoted by $P_{q}(\theta, \phi)$, of our developed EMUSIC algorithm can be calculated by
\begin{equation}
    P_{q}(\theta, \phi) = \frac{1}{ {\bf a}_{q}^{ H}\widehat{\bf U}_{q,n}\widehat{\bf U}_{q,n}^{\rm H}{\bf a}_{q} }.
    \label{eq:spectral}
\end{equation}
By searching the spectrum peaks in both $(\theta,\phi)$, the estimations of $(\theta_{g},\phi_{g})$, denoted by $(\hat{\theta}_{qg},\hat{\phi}_{qg})$, for the given $q$-th subcarrier can be acquired according to Eq.~\eqref{eq:spectral}.

\emph{Proposition 1}: When the coefficients $\rho$ and $\nu$ respectively satisfy $\rho \in (0,1]$ and $\nu > 0$, $G$ targets of developed EMUSIC algorithm can be correctly detected under unknown number of targets.
\begin{proof}
Proof see in Appendix~\ref{Appendi:EMUSIC}.
\end{proof}

{\bf 2) $R_{g}$, $v$, and $\phi_{g}$ estimation in the frequency domain:}
According to Eq.~\eqref{eq:hqisq}, the corresponding vector, denoted by ${\bf h}_{i_{s}} \in \mathcal{C}^{N_{f}\times 1}$, of ${\bf H}_{s}$ consisting of the information about $\phi_{g}$, $R_{g}$, and $v$ for the $i_{s}$-th OAM mode can be expressed by
\begin{equation}
    {\bf h}_{i_{s}} = {\bf A}_{i_{s}} {\bf \chi} +{\bf n}_{i_{s}},
    \label{eq:his}
\end{equation}
where ${\bf n}_{i_{s}} \in \mathcal{C}^{N_{f}\times 1}$ is the noise vector for the $i_{s}$-th OAM mode and the entry in the $i_{s}$-th row and the $g$-th column of ${\bf A}_{i_{s}}$ is $A_{q,i_{s},g} e^{-j 2 \pi f_{0} \tau_{g}} e^{j2\pi f_{d} i_{s} T_{0}}e^{-j 2\pi q \Delta f \tau_{g}}$. The matrix ${\bf A}_{i_{s}}$ contains the elevation angle, relative range, and velocity information of all reflect points on the $l_{i_{s}}$ OAM beam.
Similar to the analysis of the estimates regarding $(\theta_{g},\phi_{g})$ in the OAM domain and according to Eq.~\eqref{eq:his}, we can derive the estimates of $(\phi_{g},R_{g})$, denoted by $(\hat{\phi}_{i_{s}g},\hat{R}_{i_{s}g})$, for the given $l_{i_{s}}$ OAM beam by searching the spectrum peaks of the spectral function, denoted by $P_{i_{s}}(\phi, R)$, in both elevation angle and relative range $(\phi, R)$ domains as follows:
\begin{eqnarray}
   P_{i_{s}}(\phi, R)=\frac{1}{{\bf b}_{i_{s}}^{ H} \widehat{\bf U}_{i_{s},n} \widehat{\bf U}_{i_{s},n}^{ H} {\bf b}_{i_{s}}},
\end{eqnarray}
where $\widehat{\bf U}_{i_{s},n}$ is the EMUSIC-based noise subspace composed of the $(N_{f}-\widehat{G})$ eigenvectors corresponding to the $(N_{f}-\widehat{G})$ smallest eigenvalues of the covariance matrix regarding ${\bf h}_{i_{s}}$ and ${\bf b}_{i_{s}}=\left[b_{1l_{i_{s}}},\cdots,b_{ql_{i_{s}}},\cdots,b_{N_{f}l_{i_{s}}}\right]^{ T}$ is the reconstructed direction vector given by
\begin{eqnarray}
    b_{ql_{i_{s}}} &=& \frac{ \lambda_{q} \!  e^{-j\frac{2\pi}{\lambda_{q}}\left(\sqrt{R^{2}+r_{t}^{2}}+\sqrt{R^{2}+r_{r}^{2}}\right)}
  J_{0}\!\!\left(\!\!\frac{2\pi r_{r}R\sin\phi}{\lambda_{q} \sqrt{R^{2}+r_{r}^{2}}}\!\!\right)}{\sqrt{ (R^2+r_{t}^2 ) (R^2+r_{r}^2 ) }}
  \nonumber\\
  && J_{l_{i_{s}}}\!\!\!\left(\!\!\frac{ 2\pi r_{t} R \sin\phi } {\lambda_{q} \sqrt{R^{2}+r_{t}^{2}}}\!\! \right) e^{-j2\pi \frac{2R}{c} (f_{0}+q \Delta f) }.
\end{eqnarray}
The estimation of relative velocity $v$ also can be obtained with the EMUSIC algorithm.

Therefore, based on the estimations of the azimuth angle, the elevation angle, and relative ranges analyzed above, we have the estimates, denoted by $\hat{\theta}_{g}$, $\hat{\phi}_{g}$, and $\hat{R}_{g}$, for the $g$-th reflect point as follows:
\begin{equation}
    \begin{cases}
        \hat{\theta}_{g}=\frac{1}{N_{f}} \sum\limits_{q=1}^{N_{f}} \hat{\theta}_{qg};
       \\
        \hat{\phi}_{g} = \frac{1}{N_{t}N_{f}}\left(\sum\limits_{q=1}^{N_{f}} \hat{\phi}_{qg} + \sum\limits_{i_{s}=1}^{N_{t}} \hat{\phi}_{i_{s}g}\right);
        \\
         \hat{R}_{g}= \frac{1}{N_{t}}\sum\limits_{i_{s}=1}^{N_{t}} \hat{R}_{i_{s}g}.
    \end{cases}
    \label{eq:theta_es}
\end{equation}



Based on Eq.\eqref{eq:theta_es}, the position of the jammer is estimated. The accuracy of estimation depends on the SSNR, number of sensing OAM modes $N_{t}$, and the number of subcarriers $N_{f}$.

\vspace{5pt}
\subsection{Multiuser Signal Extraction and Mode Allocation}
With the estimated position of the jammer with $N_{\rm J}$ antennas in Section~\ref{subsec:position}, the relative locations between the smart jammer and the $K$ legitimate users can be estimated. Thus, denoting ${\bf H}_{{\rm J},kq}$ the channel from the jammer to the $k$-th legitimate user at the $q$-th subcarrier, we have the $u$-th $1 \leq u \leq N_{t}$ row and the $n_{j}$-th $(1 \leq n_{j} \leq N_{\rm J})$ column of ${\bf H}_{{\rm J},kq}$ as follows:
 \begin{equation}
    h_{{\rm J},un_{j},kq}=\frac{\beta_{{\rm J},k} \lambda_{q}}{4\pi d_{un_{j},k}} e^{j\frac{2\pi d_{un_{j},k} }{\lambda_{q}}},
\end{equation}
where $\beta_{{\rm J},k}$ is the attenuation constant and $d_{un_{j},k}$ is the jamming distance from the jammer to the $u$-th receive array through the LoS communication link for the $k$-th legitimate user. 
The perfect communication CSI is assumed to be known by the ISAC transmitter and $K$ legitimate OAM users. Thereby, we have the channel gain, denoted by ${\bf H}_{kq}$, from the OAM transmitter to the $k$-th legitimate user through the LoS communication link at the $q$-th subcarrier, where ${\bf H}_{kq}$ consists of the entries $h_{kq,un}$ in the $u$-th row and the $n$-th column. $h_{kq,un}$ can be obtained by replacing $\beta_{{\rm J},k}$ by $\beta_{k}$ and $d_{un_{j},k}$ by $d_{k,un}$, where $d_{k,un}$ denotes the communication distance from the $n$-th transmit array to the $u$-th receive array for the $k$-th legitimate user.

In most practical scenarios, the transmit UCA and the receive UCAs are misaligned, thus resulting in the non-circular of ${\bf H}_{kq}$. This indicates that inter-user and inter-mode interference mitigation is limited when only the FFT algorithm is used at the users. To further mitigate them, the joint transmit and receive beamforming for each legitimate user is required before decomposing OAM signals. 
Therefore, the received decomposing OAM signal, denoted by ${\bf y}_{kq}$, for the $k$-th legitimate user at the $q$-th subcarrier is calculated by
\begin{align}
  {\bf y}_{kq} &= {\bf F }{\bf W}_{{\rm rx},kq}^{ H}{\bf H}_{kq} {\bf W}_{{\rm tx},kq}{\bf x}_{q,{ c}} \! +\! {\bf F }{\bf W}_{{\rm rx},kq}^{ H}\left({\bf H}_{{\rm J},kq} {\bf x}_{{\rm J}}
  \!+\! {\bf n}_{k,q}\right)
  \nonumber\\
  &=\underbrace{{\bf F }{\bf W}_{{\rm rx},kq}^{ H}{\bf H}_{kq} {\bf W}_{{\rm tx},kq}{\bf F}^{H}{\bf P}_{qk}{\bf s}_{q,k}}_{{\rm legitimat \ signal}}+\underbrace{{\bf F }{\bf W}_{{\rm rx},kq}^{ H}{\bf H}_{{\rm J},kq} {\bf x}_{{\rm J}}}_{{\rm jamming}}
  \nonumber\\
  &\!+\!\underbrace{\sum_{i=1, i\neq k}^{K}\!{\bf F }{\bf W}_{{\rm rx},kq}^{ H}\!{\bf H}_{kq} \!{\bf W}_{{\rm tx},kq}{\bf F}^{ H}{\bf P}_{kq}{\bf s}_{q,i}}_{{\rm inter-user \ interference}}+{\bf F }{\bf W}_{{\rm rx},kq}^{ H}{\bf n}_{kq},
   \label{eq:decom}
\end{align}
where ${\bf W}_{{\rm rx},kq}$, ${\bf W}_{{\rm tx},kq}$, and ${\bf n}_{kq}$ denote the receive beamformer, transmit beamformer, and the received AWGN with zero mean and variance $\sigma_{kq}^{2}{ \bf I}_{N}$, respectively, for the $k$-th legitimate receiver at the $q$-th subcarrier. Also, ${\bf x}_{{\rm J}}$ is supposed to be the jamming. As shown in Eq.~\eqref{eq:decom}, the $k$-th legitimate user suffers from jamming and inter-user interference in addition to the received legitimate signals.

Owing to the pre-shared index information between the transmitter and $K$ legitimate users, the indices corresponding to OAM modes $\mathcal{L}_{k}$ can be easily obtained. We use the ${\bf e}_{i_{k}}$ to acquire the expected output OAM decomposed signals, denoted by $y_{qi_{k}}$, for the $i_{k}$-th OAM mode concerning the $k$-th legitimate user as follows:
\begin{eqnarray}
    y_{q i_{k}} = {\bf e}_{i_{k}}^{ H}{ \bf y}_{kq}.
    \label{eq:rik}
\end{eqnarray}
Thus, the received SINR, denoted by $\gamma_{qi_{k}}$, for the $i_{k}$-th OAM mode at the $q$-th subcarrier can be derived as follows:
\begin{equation}
  \gamma_{qi_{k}}=\frac{P_{qi_{k}}\left|{\bf f}_{i_{k}}^{H}{\bf W}_{{\rm rx},kq}^{ H}{\bf H}_{kq} {\bf W}_{{\rm tx},kq}{\bf f}_{i_{k}}\right|^{2}}{\sigma_{qi_{k}}^{2}+\Sigma_{qi_{k}}},
  \label{eq:sinr}
\end{equation}
where $\sigma_{qi_{k}}^{2} = \sigma_{kq}^2 | {\bf f}_{i_{k}}^{H}{\bf W}_{{\rm rx},kq}^{ H}|^{2}$ and $\Sigma_{qi_{k}}$ contains the inter-mode interference, inter-user interference and jamming is calculated by
\begin{align}
  \Sigma_{qi_{k}} =&  \sum_{\substack{j_{k} =1 \\ j_{k} \neq i_{k}}}^{N_{k}} P_{qj_{k}}\left|{\bf f}_{i_{k}}^{H}{\bf W}_{{\rm rx},kq}^{ H}{\bf H}_{kq} {\bf W}_{{\rm tx},kq}{\bf f}_{j_{k}}\right|^2
  \nonumber\\
 & + \sum_{\substack{i=1 \\ i\neq k}}^{K} \sum_{i_{i}=1}^{N_{i}} P_{qi_{i}} \left|{\bf f}_{i_{k}}^{H}{\bf W}_{{\rm rx},kq}^{ H}{\bf H}_{kq} {\bf W}_{{\rm tx},kq}{\bf f}_{i_{i}}\right|^2
  \nonumber\\
  & + \left|{\bf f}_{i_{k}}^{H}{\bf W}_{{\rm rx},kq}^{ H}{\bf H}_{{\rm J},kq}{\bf x}_{\rm J} \right|^2.
   \label{eq:interference}
\end{align}

\vspace{10pt}
\section{Maximizing ASR for Proposed ISAC Systems}\label{sec:capa}

\subsection{Problem Formulation}

The mutual information between the transmit signals ${\bf s}_{q,k}$ and the legitimate receive signals $ y_{qi_{k}}$ for the $K$ users can be expressed as follows~\cite{7933242}:
 \begin{equation}
   \mathcal{I}({\bf s}_{q,k}, y_{qi_{k}})=\mathcal{I}\left({\bf s}_{q,k}, y_{qi_{k}}|{\bf x}_{q,{ c}}\right)+\mathcal{I} \left({{\bf x}_{q,{ c}}},y_{qi_{k}}\right),
   \label{eq:mutual}
 \end{equation}
where
  \begin{equation}
   \mathcal{I}\left({\bf s}_{q,k}, y_{qi_{k}}|{\bf x}_{q,{ c}}\right)=\frac{1}{N_{f}}\sum_{q=1}^{N_{f}}\sum_{k=1}^{K}\sum_{i_{k}=1}^{N_{k}}\log_{2}\left(1+\gamma_{q i_{k}}\right)
 \end{equation}
represents the signal information and $\mathcal{I}\left({{\bf x}_{q,{ c}}},y_{qi_{k}}\right)$ represents the index information for communications.

Considering the $N_{t}$ OAM modes simultaneously for sensing and communications in the proposed ISAC system, we can calculate the number of OAM-mode combinations, denoted by $C_{1}$, for $1$-th OFDM symbol duration as follows:
\begin{equation}
    C_{1}=2^{\left\lfloor \log_{2}\left[ N_{t} \binom{N_{t}-1}{N_{1}}\binom{N_{t}-1-N_{1}}{N_{2}}\cdots \binom{N_{t}-1-N_{1}\cdots -N_{K-1}}{N_{K-1}} \right]\right\rfloor}.\label{eq:C1}
\end{equation}
Thus, we have $\mathcal{I}\left({{\bf x}_{q,c}},y_{qi_{k}}\right)$ within the 1-th OFDM symbol duration as follows:
\begin{equation}
 \mathcal{I}\left({{\bf x}_{q,c}},y_{qi_{k}}\right) \leq \log_{2}C_{1}.
\end{equation}
Next, considering that all the OAM modes are used for sensing during $T_{s}$, we can calculate the number of OAM-mode combinations, denoted by $(C_{2}, C_{3},\cdots, C_{N_{t}})$, for the $(2,3,\cdots, N_{t})$-th OFDM symbol duration, respectively, as follows:
\begin{equation}
    \!\!\!\!\!\begin{cases}
        \!C_{2}\!=\!2^{\!\left\lfloor \!\log_{2}\left[ (N_{t}-\!1) \binom{N_{t}-1}{N_{1}}\binom{N-1-N_{1}}{N_{2}}\cdots \binom{N_{t}-\!1-N_{1}\cdots -N_{K-\!1}}{N_{K-\!1}} \!\right]\!\right\rfloor};
        \\
        \!C_{3}\!=\!2^{\!\left\lfloor \!\log_{2}\left[ (N_{t}-\!2) \binom{N_{t}-1}{N_{1}}\binom{N-1-N_{1}}{N_{2}}\cdots \binom{N_{t}\!-1-N_{1}\cdots -N_{K-\!1}}{N_{K-\!1}} \!\right]\!\right\rfloor};
        \\
        \vdots ~~~~~~~~~~~\vdots
        \\
       \! C_{N_{t}}\!=\!2^{\!\left\lfloor \!\log_{2}\left[  \binom{N_{t}-\!1}{N_{1}}\binom{N_{t}-1-N_{1}}{N_{2}}\cdots \binom{N_{t}-\!1-N_{1}\cdots -N_{K-\!1}}{N_{K-\!1}} \!\right]\!\right\rfloor}.
    \end{cases}
\end{equation}
Therefore, the upper bound of ASR, denoted by $C_{{\rm ISAC}}$,  corresponding to all legitimate receivers for the proposed ISAC system under malicious jamming within the sensing duration $T_{s}$ can be expressed as follows:
 \begin{equation}
  C_{\rm ISAC}\!\!=\!\!\frac{1}{N_{f}}\!\!\left[\!\sum_{q=1}^{N_{f}}\sum_{k=1}^{K}\!\sum_{i_{k}=1}^{N_{k}}\!\!\log_{2} \left(1\!+\!\!\gamma_{qi_{k}}\!\right)\!+\!\log_{2}\left(\!C_{1}C_{1}\cdots C_{N_{t}}\!\right)\!\right].
  \label{eq:capacity}
 \end{equation}

To maximize the ASR in Eq.~\eqref{eq:capacity}, which is subject to the Tx-Rx beamforming and the transmit power constraints while guaranteeing the position estimation accuracy in OAM sensing, the optimization problem can be formulated as follows:
\begin{subequations}
\begin{align}	
		\textbf{P1}: ~~~~&\underset{\substack{{\bf P}_{kq}, \{{\bf W}_{{\rm tx},kq}\}, \{{\bf W}_{{\rm rx},kq}\}}}{\text{Maximize}} \quad C_{\rm ISAC} \label{eq:P1_ob}
\\
		&\mathrm{s.t}.~~~ 1)~  \text{Eq.~\eqref{eq:power_sum}}, P_{q i_{k}} \geq \bar{P},\quad \forall q,i_{k},i_{s}; \label{eq:cons_power}
\\
		&~~~~~~~2)~\parallel [{\bf W}_{{{\rm tx},kq}}]_{n} \parallel^2 \leq 1,\quad \forall k,n,q; \label{eq:cons_Wtx}
\\
		&~~~~~~~3)~\parallel [{\bf W}_{{\rm rx},kq}]_{n} \parallel^2  = 1,\quad  \forall k,n,q ; \label{eq:cons_Wrx}
\\
        &~~~~~~~4)~\gamma_{qi_{s}} \geq \gamma_{s}, \label{eq:cons_sensing}
\end{align}	
\end{subequations}
where $\gamma_{s}$ is the minimum SSNR constraint for sensing accuracy, $\gamma_{qi_{s}}$ denotes the SSNR for the $l_{i_{s}}$ OAM mode, and $\bar{P}$ represents the allocated power threshold. ${\bf W}_{{\rm tx},kq} $ and ${\bf W}_{{\rm rx},kq}$ respectively obey the normalized power constraints~\cite{9511285}.

As shown in Eq.~\eqref{eq:sinr}, Problem $\textbf{P1}$ is a non-convex optimization problem, where the object function is non-concave over the transmit beamforming matrix ${\bf W}_{{\rm tx},kq}$, the receive beamforming matrix ${\bf W}_{{\rm rx},kq}$, and power ${\bf P}_{k}$, thus resulting in the difficulty to directly obtain the optimal power allocation and beamformers for all users.
Aiming at maximizing the ASR of proposed ISAC system under hostile jamming, we develop the weighted mean-square-error (MSE)-based AO algorithm to obtain the effective solutions of Problem $\textbf{P1}$.

\subsection{Maximizing The ASR With MSE-AO Algorithm}

Since the term $\log_{2}\left(\!C_{1}C_{1}\cdots C_{N_{t}}\!\right)$ is independent of ${\bf W}_{{{\rm tx},kq}}$, ${\bf W}_{{\rm rx},kq}$, ${\bf P}_{kq}$, and ${\bf P}_{sq}$, it can be ignored in Problem $\textbf{P1}$. To make the non-convex Problem $\textbf{P1}$ more tractable, we introduce the following lemma~\cite{5756489}.

{\bf Lemma 1}: When the auxiliary variable $w_{qi_{k}}$ satisfies $w_{qi_{k}} = \epsilon_{qi_{k}}^{-1}, \forall i_{k},q$, Problem $\textbf{P1}$ to maximize the sum rate is equivalent to minimize the sum weighted MSE subject to Tx-Rx beamforming and transmit power constraints of the ISAC systems because of the identical Karush-Kuhn-Tucher (KKT) conditions and the optimal solutions, where $\epsilon_{qi_{k}}$ is the MSE for the $i_{k}$-th OAM mode over the $q$-th subcarrier.

Based on the received signal in Eq.~\eqref{eq:rik}, we can calculate $\epsilon_{qi_{k}}$ as follows:
\begin{equation}
\begin{aligned}
 \epsilon_{qi_{k}} & =  \mathbb{E}\left[\left(y_{qi_{k}}-\sqrt{P_{q,i_{k}}} s_{qi_{k}}\right)\left(y_{qi_{k}}-\sqrt{P_{q,i_{k}}} s_{qi_{k}}\right)^{*}\right]
 \\
 &= (R_{qi_{k}}-1)^2+\Sigma_{qi_{k}}+\sigma_{kq}^2 {\bf f}_{i_{k}}^{H}{\bf W}_{{\rm rx},kq}^{ H}{\bf W}_{{\rm rx},kq}{\bf f}_{i_{k}},
\label{eq:epsilon}
\end{aligned}
\end{equation}
where
\begin{equation}
R_{qi_{k}} = \sqrt{P_{qi_{k}}} {\bf f}_{i_{k}}^{H}{\bf W}_{{\rm rx},kq}^{ H}{\bf H}_{kq} {\bf W}_{{\rm tx},kq}{\bf f}_{i_{k}}.
\end{equation}
Then, Problem $\textbf{P1}$ can be transformed into Problem $\textbf{P2}$ to minimize the sum of weighted MSE on the basis of Lemma 1 as follows:
\begin{subequations}
\begin{align}	
		\textbf{P2}: &\underset{\substack{{\bf P}_{k}, {\bf P}_{\rm s}, \{w_{qi_{k}}\},\\
\{{\bf W}_{{\rm tx},kq}\}, \{{\bf W}_{{\rm rx},kq}\}}}{\text{Maximize}}~~\sum_{q=1}^{N_{f}} \sum_{k=1}^{K} \sum_{i_{k}=1}^{N_{k}} \left(\log_{2} w_{qi_{k}} \!- \!w_{qi_{k}}\epsilon_{qi_{k}}\! +\! 1 \right)
\\
		&\mathrm{s.t}.~~~ 1)~{\rm Eqs.~\eqref{eq:cons_power} ~ - ~ \eqref{eq:cons_sensing}};
\\
		&~~~~~~~2)~w_{qi_{k}} >0, ~~ \forall i_{k},q .
\end{align}	
\end{subequations}

The weighted MSE-AO algorithm is developed to tackle Problem ${\textbf{P2}}$ which is transformed into several sub-optimal problems to optimize a variable with given other variables. Then, these subproblems are alternately solved till convergence.

\emph{1) Optimize $w_{qi_{k}}$}: Given ${\bf W}_{{\rm tx},kq}$, ${\bf W}_{{\rm rx},kq}$, ${\bf P}_{sq}$, and ${\bf P}_{kq}$, the sub-problem $\textbf{P2.1}$ with regard to $w_{qi_{k}}$ is convex and unconstrained given by
\begin{align}	
	\textbf{P2.1}: ~	&\underset{\substack{\{w_{qi_{k}}\}}}{\text{Maximize}} \quad \sum_{q=1}^{N_{f}} \sum_{k=1}^{K} \sum_{i_{k}=1}^{N_{k}} \log_{2} w_{qi_{k}} - w_{qi_{k}}\epsilon_{qi_{k}}
\end{align}	
Thus, the optimal $w_{qi_{k}}$, denoted by $w_{qi_{k}}^{\star}$, can be obtained by setting the partial derivative with respect to $w_{qi_{k}}$ to zero. Then, $w_{qi_{k}}^{\star}$ can be calculated by
\begin{equation}
w_{qi_{k}}^{\star} = ((R_{qi_{k}}-1)^2+\Sigma_{qi_{k}}+\sigma_{kq}^2 {\bf f}_{i_{k}}^{H}{\bf W}_{{\rm rx},kq}^{ H}{\bf W}_{{\rm rx},kq}{\bf f}_{i_{k}})^{-1}.
\label{eq:wqik}
\end{equation}
The computational complexity of $w_{qi_{k}}$ for the $k$-th user is $\mathcal{O} \left(N_{k}N_{t}^2\right)$ mainly due to the matrix multiplication.

\emph{2) Optimize ${\bf W}_{{\rm rx},kq}$}: When $w_{qi_{k}}$, ${\bf W}_{{\rm tx},kq}$, ${\bf P}_{sq}$, and ${\bf P}_{kq}$ are all given, ${\bf W}_{{\rm rx},kq}$ is only related to the $k$-th user at the $q$-th subcarrier. Thus, the sub-problem $\textbf{P2.2}$ corresponding to ${\bf W}_{{\rm rx},kq}$ can be formulated by
\begin{subequations}
\begin{align}	
	\textbf{P2.2}:~~	&\underset{\substack{\{ {\bf W}_{{\rm rx},kq} \}}}{\text{Minimize}} \quad  \sum_{i_{k}=1}^{N_{k}} w_{qi_{k}}\epsilon_{qi_{k}}~\label{eq:P2.2}
\\
& \mathrm{s.t}.~~~ ~\parallel [{\bf W}_{{\rm rx},kq}]_{n} \parallel^2  = 1,\quad  \forall k,n,q .
\end{align}	
\end{subequations}
Although the constraint $\parallel [{\bf W}_{{\rm rx},kq}]_{n} \parallel^2  =  1$ is non-convex, we first deal with Problem $\textbf{P2.2}$ as an unconstrained convex optimal problem and then normalize the optimal solution, denoted by ${\bf W}_{{\rm rx},kq}^{\star}$, after the whole weighted MSE-AO algorithm is converged~\cite{10159012}. By utilizing
\begin{equation}
    \frac{\partial \left({\bf Xa}\right)^{ H}{\bf C} \left({\bf Xb}\right)}{\partial {\bf X}}= {\bf CXba}^{ H} + {\bf C}^{ H}{\bf Xab}^{ H},
\end{equation}
where ${\bf C}$, ${\bf a}$, ${\bf b}$ are not functions of ${\bf X}$, the partial derivative of Eq.~\eqref{eq:P2.2} corresponding to ${\bf W}_{{\rm rx},kq}$ is calculated by
\begin{equation}
\begin{aligned}
  &\sum\limits_{i_{k}=1}^{N_{k}}\frac{\partial \left(w_{qi_{k}}\epsilon_{qi_{k}}\right)}{\partial {\bf W}_{{\rm rx},kq}}
  \\
  & =\!\!\!\sum_{i_{k}=1}^{N_{k}} \!w_{qi_{k}}\!\!\left(\!\left({\bf B}_{kq} \!+\! \sigma_{kq}^{2} {\bf I}_{N_{t}}\!\right) \!\!{\bf W}_{{\rm rx},kq} \!-\!\sqrt{P_{qi_{k}}} {\bf H}_{kq} {\bf W}_{{\rm tx},kq}\!\!\right)\!{\bf f}_{i_{k}} {\bf f}_{i_{k}}^{H},
 \end{aligned}
\end{equation}
where ${\bf B}_{kq}$ is defined by
\begin{align}
{\bf B}_{kq} =& {\bf H}_{kq} \left(\sum_{i=1 }^{K}{\bf W}_{{\rm tx},iq}{\bf F} {\bf P}_{iq}{\bf P}_{iq}^{T} {\bf F}^{H}{\bf W}_{{\rm tx},iq}^{H}\!\right){\bf H}_{kq}^{H}
 \nonumber\\
 &+{\bf H}_{{\rm J},k} {\bf x}_{\rm J} {\bf x}_{\rm J}^{H}{\bf H}_{{\rm J},k}^{H}.
\end{align}
\setcounter{equation}{60}
\begin{figure*}[!hb]
\hrulefill
  \begin{equation}
  \begin{aligned}
    {\bf P}_{kq}^{\star} = \Lambda_{kq} \left(\eta^{\star} {\bf I}_{N_{t}} + {\bf F}^{H} {\bf W}_{{\rm tx},kq}^{H} \left(\sum_{i=1}^{N_{k}} {\bf H}_{iq}^{H} {\bf W}_{{\rm rx},iq} {\bf F} {\bf W}_{iq} {\bf F}^{H}
    {\bf W}_{{\rm rx},iq}^{H} {\bf H}_{iq}\right) {\bf W}_{{\rm tx},kq} {\bf F}\right)^{-1},
  \end{aligned}
  \label{eq:Pk}
\end{equation}
\end{figure*}
Then, setting $\sum_{i_{k}=1}^{N_{k}}\frac{\partial \left(w_{qi_{k}}\epsilon_{qi_{k}}\right)}{\partial {\bf W}_{{\rm rx},kq}}={\bf 0}_{N_{t}}$, we can derive the optimal ${\bf W}_{{\rm rx},kq}^{\star}$ by
\setcounter{equation}{54}
\begin{align}
 {\bf W}_{{\rm rx},kq}^{\star} =& \left({\bf B}_{kq} + \sigma_{kq}^{2}{\bf I}_{N_{t}}\right)^{-1} {\bf H}_{kq} {\bf W}_{{\rm tx},kq} {\bf F} {\bf P}_{kq} {\bf P}_{kq}^{H} {\bf W}_{kq} {\bf F}^{H}
\nonumber\\
& \times \left({\bf F}  {\bf W}_{kq} {\bf F}^{H}\right)^{-1},
\label{eq:Wrx_op}
\end{align}
where ${\bf W}_{kq} ={\rm diag} \{0,w_{qi_{k}},\cdots,w_{qi_{N_{k}}}\}$ is the diagonal matrix composed of $w_{qi_{k}}$. It is noticed that the entries of ${\bf W}_{kq}$ are zeros except the entries corresponding to the OAM mode $l_{i_{k}}$.

The complexity of Eq.~\eqref{eq:Wrx_op} is $\mathcal{O}(N_{t}^{3})$ mainly due to the matrix inversion.

\emph{3) Optimize ${\bf W}_{{\rm tx},kq}$}: With other variables being fixed, Problem $\textbf{P2}$ is transformed into convex optimization Problem $\textbf{P2.3}$ with regard to ${\bf W}_{{\rm tx},kq}$ as follows:	
\begin{subequations}
\begin{align}	
		\textbf{P2.3}: ~~~~&\underset{\{{\bf W}_{{\rm tx},kq}\}}{\text{Minimize}}  \quad  \sum_{k=1}^{K} \sum_{i_{k}=1}^{N_{k}} w_{qi_{k}} \epsilon_{qi_{k}}
\\
		&\mathrm{s.t}.~~~ \parallel [{\bf W}_{{\rm tx},kq}]_{n} \parallel^2 \leq 1,\quad \forall n,q,k.
\end{align}	
\end{subequations}
Based on the KKT condition, the Lagrangian function of Problem $\textbf{P2.3}$ with regard to ${\bf W}_{{\rm tx},kq}$ and dual variable $\zeta_{kq} > 0$ can be given by
\begin{align}
\mathcal{L}({\bf W}_{{\rm tx},kq}, \zeta_{kq}) =  \sum_{k=1}^{K} \sum_{i_{k}=1}^{N_{k}} w_{qi_{k}} \epsilon_{qi_{k}} \!+ \! \zeta_{kq} \left( [{\bf W}_{{\rm tx},kq}]_{n} \parallel^2 \!-\! 1\right).
\end{align}
Then, the first-order derivative of $\mathcal{L}({\bf W}_{{\rm tx},kq}, \zeta_{kq})$ with respect to each ${\bf W}_{{\rm tx},kq}$ and $\zeta_{kq}$ is calculated. Consequently, ${\bf W}_{{\rm tx},kq}$ are derived by
\begin{equation}
  \begin{aligned}
    {\bf W}_{{\rm tx},kq} = & \!\left(\!\sum_{i=1}^{K} {\bf H}_{iq}^{H} {\bf W}_{{\rm rx},iq} {\bf F} {\bf W}_{iq}^{H} {\bf F}^{H} {\bf W}_{{\rm rx},iq}^{H} {\bf H}_{iq}\! + \!\zeta_{kq}{\bf I}_{N_{t}}\!\!\right)^{-1}
    \nonumber\\
    & \times {\bf H}_{kq}^{H} {\bf W}_{{\rm rx},kq} {\bf F} {\bf W}_{kq}^{H} {\bf P}_{kq}^{-1} {\bf F}^{H}
  \end{aligned}
  \label{eq:Wtxkq}
\end{equation}
with the optimality condition
\begin{equation}
 [{\bf W}_{{\rm tx},kq}]_{n} \parallel^2=1.
 \label{eq:Wtx_cond}
\end{equation}
The optimal $\zeta_{kq}^{\star}$ can be obtained with the method of bisection search by substituting Eq.~\eqref{eq:Wtxkq} with Eq.~\eqref{eq:Wtx_cond}. Thereby, the optimal solution ${\bf W}_{{\rm tx},kq}$, denoted by ${\bf W}_{{\rm tx},kq}^{\star}$, is calculated by replacing $\zeta_{kq}$ with $\zeta_{kq}^{\star}$ in Eq.~\eqref{eq:Wtxkq} as follows:
\begin{equation}
  \begin{aligned}
    {\bf W}_{{\rm tx},kq}^{\star} = & \!\left(\!\sum_{i=1}^{K} {\bf H}_{iq}^{H} {\bf W}_{{\rm rx},iq} {\bf F} {\bf W}_{iq}^{H} {\bf F}^{H} {\bf W}_{{\rm rx},iq}^{H} {\bf H}_{iq}\! + \!\zeta_{kq}^{\star} {\bf I}_{N_{t}}\!\!\right)^{-1}
    \nonumber\\
    & \times {\bf H}_{kq}^{H} {\bf W}_{{\rm rx},kq} {\bf F} {\bf W}_{kq}^{H} {\bf P}_{kq}^{-1} {\bf F}^{H}.
  \end{aligned}
  \label{eq:Wtxkq_op}
\end{equation}
Mainly due to the matrix multiplication and the bisection search for $\zeta_{kq}$ with the number of iteration $I_{1}$, the complexity of ${\bf W}_{{\rm tx},kq}$ is $\mathcal{O}\left(I_{1}N_{t}^3\right)$.

\emph{4) Optimize ${\bf P}_{kq}$}: With given $\gamma_{s}$, we can obtain ${\bf P}_{sq}$ satisfying the condition $\gamma_{qi_{s}} = \gamma_{s}$. Then, fixed ${\bf W}_{{\rm tx},kq}$, ${\bf W}_{{\rm rx},kq}$, and $w_{qi_{k}}$, the convex optimization sub-problem $\textbf{P2.4}$ with respect to ${\bf P}_{kq}$ is given by
 \begin{subequations}
 \begin{align}	
	\textbf{P2.4}:~~	&\underset{\substack{\{ {\bf P}_{k} \}}}{\text{Minimize}} \quad  \sum_{q=1}^{N_{f}} \sum_{k=1}^{K} \sum_{i_{k}=1}^{N_{k}} w_{qi_{k}}\epsilon_{qi_{k}}~\label{eq:P2.2}
\\
& \mathrm{s.t}.~~~ ~ 1)~ {\rm Tr}\left(\sum_{q=1}^{N_{f}}{\bf P}_{sq}{\bf P}_{sq}^{T}+\sum_{k=1}^{K}\sum_{q=1}^{N_{f}}{\bf P}_{kq}{\bf P}_{kq}^{T}\right) \leq P_{t};
\\
& ~~~~~~ ~~2)~ P_{qi_{k}} \geq \bar{P}. \quad  \forall q,i_{k} .
\end{align}
\end{subequations}

Similar to the design of block ${\bf W}_{{\rm tx},kq}$, the Lagrangian function of Problem $\textbf{P2.3}$ with regard to ${\bf W}_{{\rm tx},kq}$ according to the KKT condition is formulated by
\begin{align}
\widetilde{\mathcal{L}}({\bf P}_{kq}, \xi_{kq},\eta) = & \sum_{q=1}^{N_{f}} \sum_{k=1}^{K} \sum_{i_{k}=1}^{N_{k}} w_{qi_{k}}\epsilon_{qi_{k}} + \xi_{kq}(P_{qi_{k}} -\bar{P} )
\nonumber\\
&  +\!\! \eta\!\! \left({\rm Tr}\!\!\left(\!\sum_{q=1}^{N_{f}}\!\left(\!\!{\bf P}_{sq}{\bf P}_{sq}^{T}\!+\!\!\sum_{k=1}^{K}\!{\bf P}_{kq}{\bf P}_{kq}^{T}\!\!\right)\!\!\right) \!-\! P_{t}\!\!\right),
\end{align}
where $\eta >0 $ and $\xi_{kq}>0$ are dual variables. Taking the first-order derivative of $\widetilde{\mathcal{L}}({\bf P}_{kq}, \xi_{kq},\eta)$ corresponding to ${\bf P}_{kq}$, $\eta$, and $\xi_{kq}$ and then setting them to zeros, we thus can obtain the optimal solutions ${\bf P}_{kq}^{\star}$ as shown in Eq.~\eqref{eq:Pk},
where the optimal solution $\eta^{\star}$ can be obtained by substituting Eq.~\eqref{eq:Pk} into
\setcounter{equation}{61}
\begin{equation}
{\rm Tr}\left(\sum_{q=1}^{N_{f}}{\bf P}_{sq}{\bf P}_{sq}^{T}+\sum_{k=1}^{K}\sum_{q=1}^{N_{f}}{\bf P}_{kq}{\bf P}_{kq}^{T}\right) = P_{t}
\end{equation}
with the bisection search,
 and $\Lambda_{kq}$ is a diagonal matrix with the diagonal entry $\Lambda_{kq,n}$ defined as
\begin{equation}
  \Lambda_{kq,n} = \left\{
  \begin{array}{lll}
     {\bf f}_{i_{k}} {\bf W}_{{\rm rx},kq}^{H} {\bf H}_{kq} {\bf W}_{{\rm tx},kq} {\bf f}_{i_{k}},  &{\text{if}~ n = l_{i_{k}} + \left\lfloor\frac{N_{t}}{2}\right\rfloor};
    \\
    0,  &{\text{otherwise}}.
  \end{array}
  \right.
\end{equation}
Updating ${\bf P}_{kq}$ in Eq.~\eqref{eq:Pk} requires $\mathcal{O}\left(I_{2} N_{k}N_{t}^2\right)$ mainly caused by the matrix multiplication, where $I_{2}$ represents the number of iterations to search $\eta$.

\emph{5) Summary}:
Initializing appropriate value of the optimized variables ${\bf W}_{{\rm tx},kq}$, ${\bf W}_{{\rm rx},kq}$, and ${\bf P}_{kq}$ under the constraints in Problem $\textbf{P1}$ and alternatively updating the optimal value of these variables, the optimal solutions can be obtained with the above-mentioned weighted MSE-AO algorithm. After the convergence of ASR in the whole iteration, ${\bf W}_{{\rm rx},kq}$ is normalized to satisfy the receive power constraint as follows:
\begin{equation}
 [{\bf W}_{{\rm rx},k}]_{n}^{\star} = \frac{ [{\bf W}_{{\rm rx},k}]_{n}^{\star}}{ \parallel  [{\bf W}_{{\rm rx},k}]_{n}^{\star} \parallel^2 }.
 \label{Eq:WRX_nor}
\end{equation}
We summarize our proposed weighted MSE-AO algorithm to obtain the maximum sum rate of our proposed ISAC in ${\textbf{Algorithm 1}}$.
\renewcommand\arraystretch{2.5}  
\begin{table*}[b]
\begin{center}
\vspace{-0pt}
\caption{Complexity comparison of proposed ISAC and traditional MIMO-ISAC schemes}
\label{table:complexity}
\vspace{0pt}
\begin{tabular}{ m{3.8cm}|  m{12cm}}
\hline
\rowcolor[gray]{0.85}
Scheme & Complexity\\
  \hline
  \cellcolor[gray]{0.92}\multirow{2}{*}{}  & \cellcolor[gray]{0.92} Joint Tx-Rx beamformer and power allocation:  $\mathcal {O}\left(\!N_{f}I_{3}\left(\!\sum\limits_{k=1}^{K}(1\!+\!I_{2})N_{k}N_{t}^2\!+\!K(I_{1}\!+\!1)N_{t}^3\right)\right)$
  \\
  \cline{2-2}
  \cellcolor[gray]{0.92} \multirow{-2}{*}{Proposed scheme} &  \cellcolor[gray]{0.98} Signal detection: $\mathcal{O} \left(N_{f}\sum\limits_{k=1}^{K}N_{k}N_{t}^2\right)$
 \\
  \hline
  \cellcolor[gray]{0.92}\multirow{2}{*}{} & \cellcolor[gray]{0.92} Joint Tx-Rx beamformer and power allocation:  $\mathcal {O}\left(\!N_{f}KI_{3}\left(N_{t}^3+N_{t}^3+ I_{1}N_{t}^3+ I_{2}N_{t}^3\right)\right)$
  \\
  \cline{2-2}
  \cellcolor[gray]{0.92} \multirow{-2}{*}{Traditional MIMO-ISAC schemes} &  \cellcolor[gray]{0.98} Signal detection: $\mathcal{O} \left(N_{f} K N_{t}^3\right)$
\end{tabular}
\end{center}
\vspace{-0pt}
\end{table*}

\begin{algorithm}
\label{Code:1}
\caption{\textbf{: Proposed MMSE-Based Joint Tx-Rx beamformer and Power Allocation Design}}
 {\bf Input:} $N_{t}$, $K$, ${\bf H}_{{\rm J},k}$, ${\bf H}_{kq}$, $P_{t}$, $\bar{P}$, ${\bf F}$, and $\gamma_{\rm s}$.
 \\
 {\bf Output:} $w_{qi_{k}}^{\star}$, ${\bf W}_{{\rm rx},kq}^{*}$, ${\bf W}_{{\rm tx},kq}^{*}$, and ${\bf P}_{kq}^{\star}$.

 \begin{algorithmic}[1]
\STATE Initialize ${\bf W}_{{\rm rx},k}$, ${\bf W}_{{\rm tx},kq}$, and ${\bf P}_{kq}$;
\STATE \textbf{while} convergence of the objective in Eq.~\eqref{eq:P1_ob} \textbf{do}
\STATE ~~Update $w_{qi_{k}}^{\star}$ by Eq.~\eqref{eq:wqik} with given other variables;
\STATE ~~Update ${\bf W}_{{\rm rx},kq}^{*}$ by Eq.~\eqref{eq:Wrx_op} with given other variables;
\STATE ~~Update ${\bf W}_{{{\rm tx},kq}}^{*}$ by Eq.~\eqref{eq:Wtxkq_op} with given other variables;
\STATE ~~Update ${\bf P}_{kq}^{\star}$ by Eq.~\eqref{eq:Pk} with given other variables;
\STATE \textbf{end}
\STATE Normalize ${\bf W}_{{\rm rx},kq}^{*}$ by Eq.~\eqref{Eq:WRX_nor}.
\STATE Return $w_{qi_{k}}^{\star}$, ${\bf W}_{{\rm rx},kq}^{*}$, ${\bf W}_{{\rm tx},kq}^{*}$, and ${\bf P}_{kq}^{\star}$.
\end{algorithmic}
\end{algorithm}

\subsection{Complexity Analysis}
The weighted MSE of our proposed ISAC systems decreases monotonically with the method of Algorithm~\ref{Code:1}, which is lower bounded. Hence, the ASR is convergent as the number of iteration increases. The computational complexities of our proposed scheme and traditional MIMO-ISAC systems are listed in Table~\ref{table:complexity}, where $I_{3}$ represents the number of iterations. It is clear that our proposed ISAC has lower complexity than the traditional MIMO-ISAC systems due to the inverse DFT (IDFT)/DFT processing at the OAM-transmitter/receiver.

\section{Numerical Results}\label{sec:Num}
In this section, we present several numerical results to evaluate the performance of our proposed ISAC scheme against malicious jamming. Throughout the whole simulation, the parameters are set as follows: $N_{f}=16$, $\Delta f =200$ kHz, $r_{t}= 0.5$ m,  $r_{r} = 0.25$ m, $f_{0} = 2.4$ GHz, the position of the jammer as $(39 ~{\rm m}, 10^{\circ}, 50^{\circ})$, $v=3$ m/s, $\gamma_{s}$ = 20 dB, $\rho =1$, $\nu =1$, $\bar{P} = \frac{1}{N_{t}^2} \sum_{k=1}^{K} \sum_{q=1}^{N_{f}} {\bf P}_{kq} {\bf P}_{kq}^{T}$, and the radii of users as 0.5 m. Also, the UCAs between the OAM transmitter and uers are misaligned.

\begin{figure}
\centering
  \includegraphics[width=0.5\textwidth]{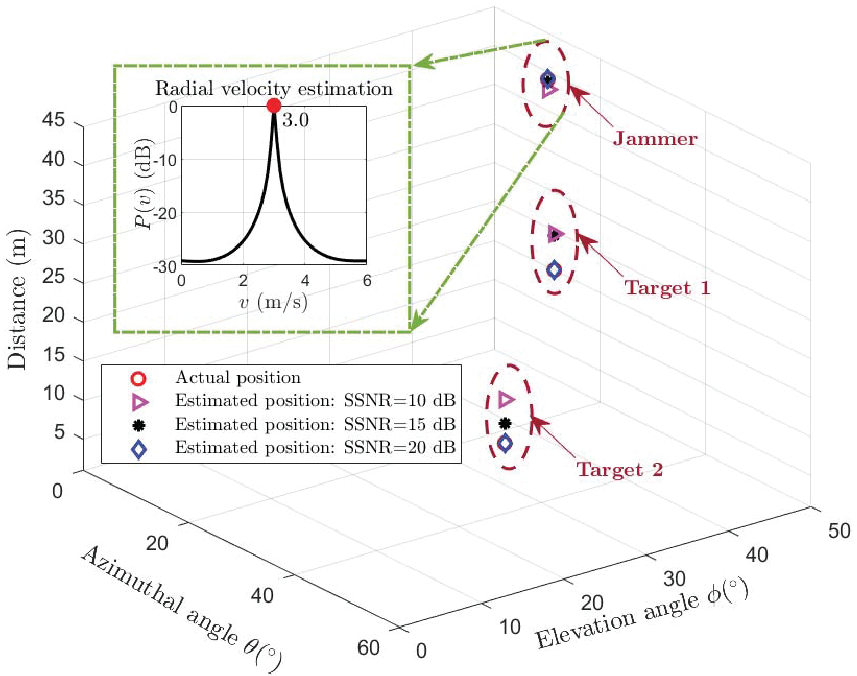}\\
  \caption{The position estimation of the jammer versus different sensing SNR.}
  \label{fig:SNR_jammer_location}
\end{figure}

Figure~\ref{fig:SNR_jammer_location} shows the target 3D position estimation of our developed EMUSIC scheme versus the SSNR, where we set the positions of targets around the jammer as $(25 ~{\rm m}, 30^{\circ}, 38^{\circ})$ and $(18 ~{\rm m}, 55^{\circ}, 16^{\circ})$, respectively. Also, we set $N_{t}=16$. It can be seen that the estimated relative 3D positions become closer to the actual positions as the sensing SNR increases, thus meaning high estimation accuracy. Also, the estimated relative radial velocity corresponding to the peaks of spectral function $P(v)$ approaches the actual value when SSNR is 20 dB. Since the targets remains stationary within the sensing duration, combining the estimated 3D position and radial velocity, the jammer can be distinguished from other targets, thus obtaining the jamming CSI by the OAM transmitter.

\begin{figure}
\centering
\subfigure[Estimation of azimuthal angles.]
{\label{fig:impact_P_azimuth} 
\includegraphics[width=0.9\linewidth]{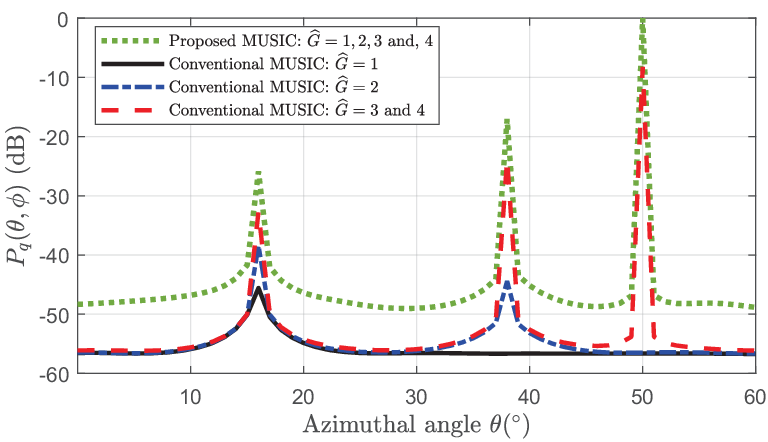}}
\subfigure[Estimation of elevation angles.]
{\label{fig:impact_P_elevation}
\includegraphics[width=0.9\linewidth]{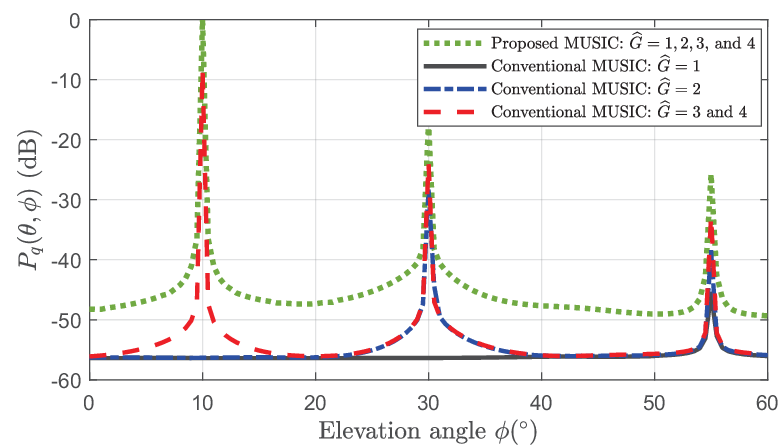}}
\caption{Estimated position comparison of our proposed and conventional MUSIC schemes.}
\label{fig:impact_P} 
\end{figure}

To verify the effectiveness of our proposed EMUSIC algorithm, we compare the spectral function results with the conventional MUSIC algorithm in Fig.~\ref{fig:impact_P}, where we set SSNR = 20 dB and $N_{t}=16$. The number of targets are assumed to be estimated as $\widehat{G} =1,2,3$ and 4. As shown in Figs.~\ref{fig:impact_P_azimuth} and~\ref{fig:impact_P_elevation}, our proposed EMUSIC algorithm can estimate the angles of all targets, while the conventional MUSIC algorithm can only acquire the angle information of the nearest $\widehat{G}$ targets corresponding to the strongest $\widehat{G}$ echo signals. By coincidence, the jammer farthest away from the OAM transmitter in the simulation, thus incorrectly estimating the location of jammer under $\widehat{G} =1$ and 2. Also, it can be found that the peaks of our proposed scheme in the target direction is higher, which can be demonstrated by Eq.~\eqref{eq:uu}. Hence, Fig.~\ref{fig:impact_P} verifies that the estimation results of our proposed EMUSIC algorithm is almost unaffected by the number of estimated targets, especially when targets are far away from the OAM transmitter.

\begin{figure}
\centering
  \includegraphics[width=0.9\linewidth]{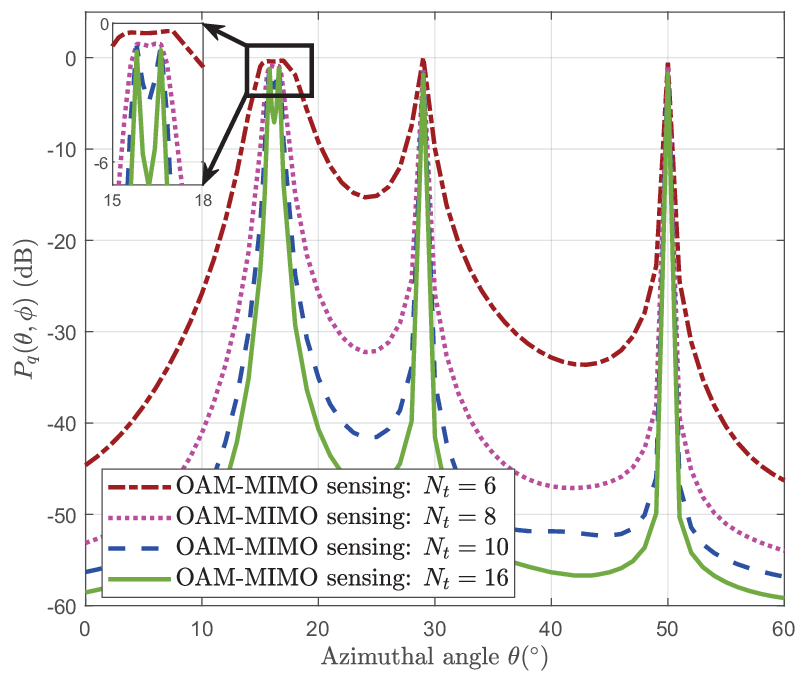}\\
  \caption{The impact of $N_{t}$ on the azimuthal resolution.}
  \label{fig:Nt_MISO_MIMO_jammer}
\end{figure}

To demonstrate the impact of the number of OAM modes on the azimuthal resolution of our proposed scheme, Fig.~\ref{fig:Nt_MISO_MIMO_jammer} depicts the values of spectral function $P_{q}(\theta,\phi)$ over the $q$-th subcarrier versus azimuthal angles, where targets have the same distances and elevation angles with but different azimuthal angles from the jammer. Thus, the azimuthal angles of targets are set as $15.8^{\circ}$, $16.6^{\circ}$, and $29^{\circ}$, respectively. The estimated azimuthal angles of the jammer and targets can be obtained by searching the peaks of $P_{q}(\theta,\phi)$. It can be observed that the estimated azimuthal angles approach the actual values with enough large $N_{t}$, where the estimated azimuthal angles for $15.8^{\circ}$ and $16.6^{\circ}$ can better illustrate this conclusion. Therefore, the conclusion is conducted that a large number of OAM modes can be utilized to achieve high azimuthal resolution for target sensing.

\begin{figure}
\centering
\subfigure[$\phi_{g}=10^{\circ}$]
{\label{fig:MIMO_vs_MISO_OAM_10} 
\includegraphics[width=0.9\linewidth]{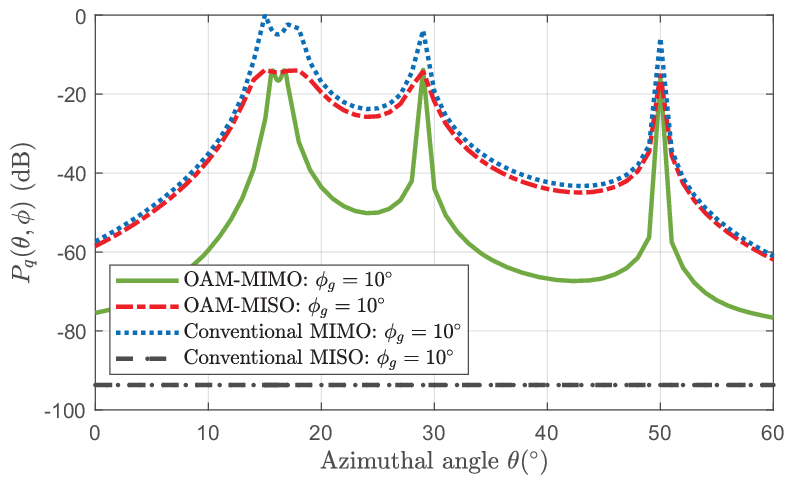}}
\subfigure[$\phi_{g}=25^{\circ}$]
{\label{fig:MIMO_vs_MISO_OAM_25}
\includegraphics[width=0.9\linewidth]{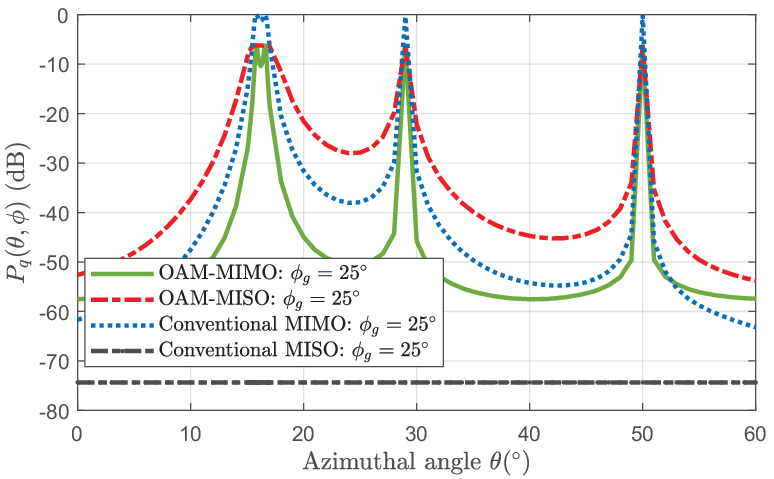}}
\caption{Comparison of estimated azimuthal angles with same elevation angles.}
\label{fig:MIMO_vs_MISO_OAM} 
\end{figure}

To verify the high azimuthal resolution of our proposed scheme, Fig.~\ref{fig:MIMO_vs_MISO_OAM} compares the spectral function $P_{q}(\theta,\phi)$ of several different UCA-based schemes with $N_{t} =16$ and SSNR=20 dB versus the azimuthal angles, where "OAM-MISO" represents the OAM based mono-static multiple-input single-output (MISO) scheme. Also, the azimuthal angle and distance settings are the same with those in Fig.~\ref{fig:Nt_MISO_MIMO_jammer}. Comparing the peaks of $P_{q}(\theta,\phi)$ in Figs.~\ref{fig:MIMO_vs_MISO_OAM_10} and~\ref{fig:MIMO_vs_MISO_OAM_25}, we can find that our proposed scheme have the highest accuracy of estimated azimuthal angles among these four schemes in the low elevation angle region. The estimated angles for OAM-MIMO, OAM-MISO, and conventional MIMO are $(15.8^{\circ},16.6^{\circ},29^{\circ},50^{\circ})$, $(15.9^{\circ},16.45^{\circ},29^{\circ},50^{\circ})$, and $(15.5^{\circ},17^{\circ},29^{\circ},50^{\circ})$, respectively. The reason is that the azimuthal resolution of our proposed scheme is determined by $e^{2j\theta l}$ as shown in Eq.~\eqref{eq:Aqisg} while the OAM-MISO sensing scheme is $e^{j\theta l}$, thus drawing the results with given SSNR and $N_{t}$. Meanwhile, the conventional MIMO for sensing mainly relies on $e^{-j\frac{2\pi}{\lambda_{q}} r_{r}\sin \phi\cos\left(\frac{2\pi m}{N_{t}} -\theta\right)}$, thus resulting in lower azimuthal resolution in the low elevation angle region. In addition, the conventional MISO scheme cannot distinguish azimuthal angles with the same elevation angles due to the lack of azimuthal angle terms. With the increase of elevation angles, the azimuthal resolution of all schemes increases. Fig.~\ref{fig:MIMO_vs_MISO_OAM} verifies the superiority of OAM-MIMO sensing. Also, OAM-MIMO in ISAC requires no orthogonal beamforming design compared with the conventional MIMO.

\begin{figure}
\centering
  \includegraphics[width=0.9\linewidth]{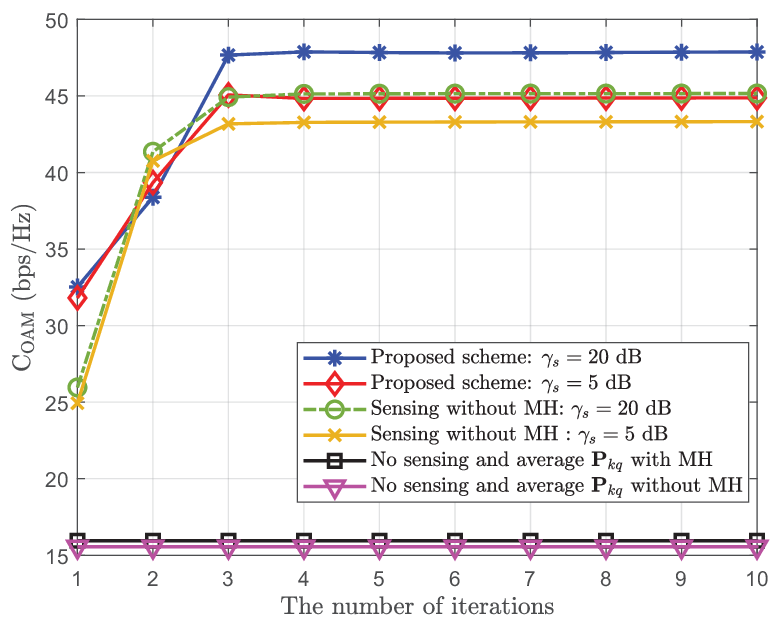}\\
  \caption{The impact of sensing results on ASR.}
  \label{fig:sensing_iteration}
\end{figure}

To demonstrate the importance of sensing results on ASR, Fig.~\ref{fig:sensing_iteration} shows the ASR of several schemes, where $N_{t} =16$, $K=2$, $N_{1} = 8$, and $N_{2} =7$. It can be seen that the ASR of our proposed scheme reaches convergence after around 3 iterations, which verifies the fast convergence. Also, the low sensing SSNR threshold causes the position estimation error, which arises the non-optimization of Tx-Rx beamforming and power allocation, thus resulting in degrading the ASR as shown in Fig.~\ref{fig:sensing_iteration}. Comparing the ASRs of our proposed scheme and the sensing scheme without MH, the conclusion is conducted that MH contributes to anti-jamming due to the change of OAM modes. Meanwhile, no sensing schemes perform worst among all the schemes in Fig.~\ref{fig:sensing_iteration}. The reason is that the Tx-Rx beamforming and power allocation cannot been optimized with unknown jamming CSI.

\begin{figure}
\centering
  \includegraphics[width=0.9\linewidth]{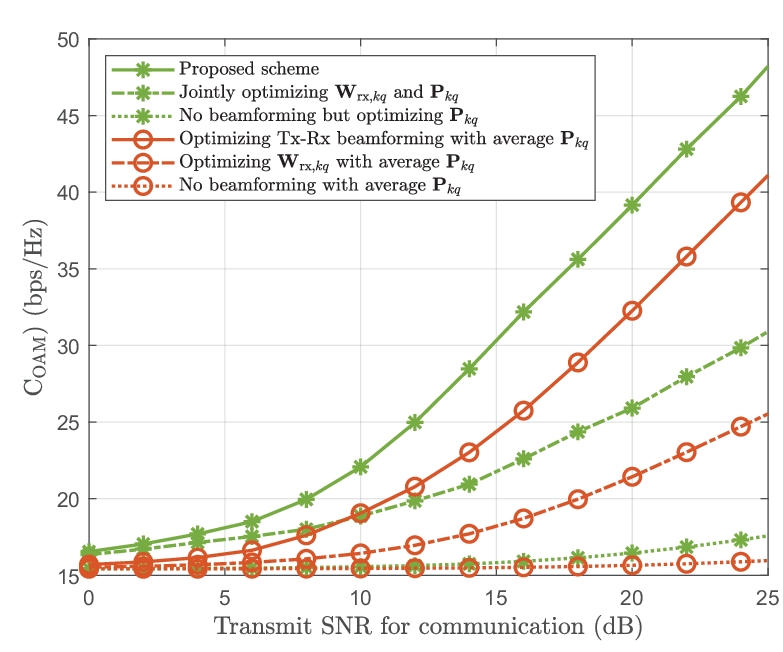}\\
  \caption{The impact of joint Tx-Rx beamforming and power allocation on anti-jamming.}
  \label{fig:SNR_beamforming_power}
\end{figure}

Figure~\ref{fig:SNR_beamforming_power} depicts the impact of joint beamforming and power allocation on ASR under hostile jamming, where $N_{t}=16$, $K=2$, $\gamma_{s}=20$ dB, $N_{1} = 8$, and $N_{2} =7$. The ASR of our proposed scheme is the highest among all the schemes in Fig.~\ref{fig:SNR_beamforming_power}. One reason is that the inter-user interference and jamming can be significantly mitigated by jointly optimizing transmit and receive beamforming under known jamming CSI. The other reason is that good channels are allocated to much power for each receivers, thus increasing the ASR.

\section{Conclusions} \label{sec:conc}

In this paper, we proposed a novel ISAC for anti-jamming with OAM system against malicious jamming, where the behavior of the jammer was unknown and dynamic. In the proposed system, the legitimate OAM-transmitter simultaneously sensed the location of the jammer and sent data to multiple legitimate receivers. The allocated OAM-modes for sensing and communication were determined by dynamic input index information. The relative CSI from the jammer to the legitimate receivers was acquired by using proposed EMUSIC algorithm in both frequency and angular domains through the continuous OAM sensing. Based on the sensing results, we designed the MMSE-based joint Tx-Rx beamforming and power allocation scheme to maximize the ASR while guaranteeing the sensing accuracy. Numerical results have verified that our proposed EMUSIC algorithm with OAM-MIMO sensing structure can achieve higher resolution than existing MIMO schemes under incorrect estimation of target number. Also, our proposed ISAC scheme with low complexity can significantly increase the ASR under hosting jamming.

\begin{appendices}
\section{Proof of Proposition 1}\label{Appendi:EMUSIC}
To obtain the range of $\rho$ and $\nu$, we have the signal subspace vectors misclassified as noise subspace vectors and the actual noise subspace vectors $\widehat{\bf U}_{q,n_{\chi}} = [\hat{\bf u}_{q,\widehat{G}+1},\cdots,\hat{\bf u}_{q,G}]$ and $\widehat{\bf U}_{q,n_{n}} = [\hat{\bf u}_{q,G+1}, \cdots, \hat{\bf u}_{q,N_{t}}]$, respectively. Similarly, we have ${\bf U}_{q,n_{\chi}} = [{\bf u}_{q,\widehat{G}+1},\cdots,{\bf u}_{q,G}]$ and ${\bf U}_{q,n_{n}} = [{\bf u}_{q,G+1},\cdots,{\bf u}_{q,N_{t}}]$.
Based on Eq.~\eqref{eq: hat_u}, we have
\begin{equation}
\frac{\widehat{\bf U}_{q,n_{\chi}} \widehat{\bf U}_{q,n_{\chi}}^{H} }{\widehat{\bf U}_{q,n_{n}} \widehat{\bf U}_{q,n_{n}}^{H} }
=
\frac{\sum\limits_{\kappa=1}^{G-\widehat{G}} \left(\frac{\rho \mu_{q,N_{t}}}{\mu_{q,\widehat{G}+\kappa}}\right)^{2\nu} {\bf u}_{q,\widehat{G}+\kappa} {\bf u}_{q,\widehat{G}+\kappa}^{H} }
{\sum\limits_{\kappa=1}^{N_{t}-G} \left(\frac{\rho \mu_{q,N_{t}}}{\mu_{q,G+\kappa}}\right)^{2\nu} {\bf u}_{q,G+\kappa} {\bf u}_{q,G+\kappa}^{H} }.
\end{equation}
Due to $\mu_{q,\widehat{G}+1} \geq \mu_{q,\widehat{G}+2}\geq \cdots \geq \mu_{q,N_{t}} >0 $, we have $\frac{\rho \mu_{q,N_{t}}}{\mu_{q,\widehat{G}+1}} \leq \frac{\rho \mu_{q,N_{t}}}{\mu_{q,\widehat{G}+2}} \leq \cdots \leq
\rho$. Thus, we have
\begin{equation}
\left\{
\begin{aligned}
& \widehat{\bf U}_{q,n_{\chi}} \widehat{\bf U}_{q,n_{\chi}}^{H} \leq \left(\frac{\rho \mu_{q,N_{t}}}{\mu_{q,G}}\right)^{2\nu} {\bf U}_{q,n_{\chi}} {\bf U}_{q,n_{\chi}}^{H};
\\
& \left(\frac{\rho \mu_{q,N_{t}}}{\mu_{q,G+1}}\right)^{2\nu} {\bf U}_{q,n_{n}}  {\bf U}_{q,n_{n}}^{H} \leq \widehat{\bf U}_{q,n_{n}} \widehat{\bf U}_{q,n_{n}}^{H}  \leq  \rho^{2\nu}  {\bf U}_{q,n_{n}}  {\bf U}_{q,n_{n}}^{H}.
\end{aligned}
\right.
\label{eq:com}
\end{equation}

Aiming at correctly detect the $(G-\widehat{G})$ targets, it requires to decrease and increase the impact of $\widehat{\bf U}_{q,n_{\chi}}$ and $\widehat{\bf U}_{q,n_{n}}$, respectively, on the spatial-spectrum function $P_{q}(\theta,\phi)$. Thus, it requires
\begin{equation}
\frac{\widehat{\bf U}_{q,n_{\chi}} \widehat{\bf U}_{q,n_{\chi}}^{H} }{\widehat{\bf U}_{q,n_{n}} \widehat{\bf U}_{q,n_{n}}^{H} } \leq \frac{{\bf U}_{q,n_{\chi}} {\bf U}_{q,n_{\chi}}^{H} }{{\bf U}_{q,n_{n}} {\bf U}_{q,n_{n}}^{H} }.
\label{eq:uu}
\end{equation}
To satisfy Eq.~\eqref{eq:uu}, it is needed that
\begin{equation}
\left\{
\begin{aligned}
& \widehat{\bf U}_{q,n_{\chi}} \widehat{\bf U}_{q,n_{\chi}}^{H} \leq   {\bf U}_{q,n_{\chi}} {\bf U}_{q,n_{\chi}}^{H};
\\
& \widehat{\bf U}_{q,n_{n}} \widehat{\bf U}_{q,n_{n}}^{H}  \geq {\bf U}_{q,n_{n}}  {\bf U}_{q,n_{n}}^{H}.
\end{aligned}
\right.
\end{equation}
According to Eq.~\eqref{eq:com}, we can obtain $ 0 <\rho \leq 1$ and $\nu > 0$. Thus, it is calculated by
\begin{equation}
\frac{\widehat{\bf U}_{q,n_{\chi}} \widehat{\bf U}_{q,n_{\chi}}^{H} }{\widehat{\bf U}_{q,n_{n}} \widehat{\bf U}_{q,n_{n}}^{H} } \leq  \left(\frac{\mu_{q,G+1}}{\mu_{q,G}}\right)^{2\nu} \frac{{\bf U}_{q,n_{\chi}} {\bf U}_{q,n_{\chi}}^{H} }{{\bf U}_{q,n_{n}} {\bf U}_{q,n_{n}}^{H} } \leq \frac{{\bf U}_{q,n_{\chi}} {\bf U}_{q,n_{\chi}}^{H} }{{\bf U}_{q,n_{n}} {\bf U}_{q,n_{n}}^{H} }.
\label{eq:uu}
\end{equation}

\end{appendices}


\bibliographystyle{IEEEbib}
\bibliography{References}

\end{document}